\documentclass[twocolumn,aps,preprintnumbers,showpacs,superscriptaddress,nofootinbib]{revtex4-1} 
\usepackage{latexsym}
\usepackage{amssymb}
\usepackage{amsmath}
\usepackage{amsthm}
\usepackage{amsfonts}
\usepackage{ifthen}
\usepackage{revsymb}
\usepackage[caption=false]{subfig}
\usepackage{yfonts}
\usepackage{graphicx}
\usepackage{algpseudocode}
\usepackage{complexity}
\usepackage{enumerate}

\usepackage[colorlinks = true, bookmarks = false]{hyperref}

\usepackage{xcolor}
\definecolor{darkred}  {rgb}{0.5,0,0}
\definecolor{darkblue} {rgb}{0,0,0.5}
\definecolor{darkgreen}{rgb}{0,0.5,0}

\hypersetup{
  urlcolor   = blue,         
  linkcolor  = darkblue,     
  citecolor  = darkgreen,    
  filecolor  = darkred       
}

\newcommand{\be}{\begin{equation}}
\newcommand{\ee}{\end{equation}}
\newcommand{\ba}{\begin{array}}
\newcommand{\ea}{\end{array}}
\newcommand{\bea}{\begin{eqnarray}}
\newcommand{\eea}{\end{eqnarray}}

\newcommand{\calA}{{\cal A }}

\newcommand{\calL}{{\cal L }}

\newcommand{\calP}{{\cal P }}

\newcommand{\calN}{{\cal N }}
\newcommand{\ZZ}{\mathbb{Z}}

\newcommand{\FF}{\mathbb{F}}

\newcommand*{\cN}{\mathcal{N}}

\newtheorem{lemma}{Lemma}
\newtheorem{prop}{Proposition}

\newtheorem{fact}{Fact}

\begin{document}

\title{Correcting coherent errors with  surface codes}

\author{Sergey \surname{Bravyi}}
\affiliation{IBM T.J. Watson Research Center}
\author{Matthias \surname{Englbrecht}}
\affiliation{Institute for Advanced Study \& Zentrum Mathematik,
Technical University of Munich}
\author{Robert \surname{K{\"o}nig}}
\affiliation{Institute for Advanced Study \& Zentrum Mathematik,
Technical University of Munich}
\author{Nolan \surname{Peard}}
\affiliation{Institute for Advanced Study \& Zentrum Mathematik,
Technical University of Munich}
\affiliation{ Department of Physics, Massachusetts Institute of Technology}

\date{\today}

\begin{abstract}
We study how well  topological quantum codes can tolerate coherent noise
caused by systematic unitary errors such as unwanted  $Z$-rotations.
Our main result is an efficient algorithm for simulating 
quantum  error correction protocols based on the 2D surface code
in the presence of coherent errors. 
The algorithm has runtime $O(n^2)$,  where $n$ is the number of physical qubits.
It allows us to simulate systems with more than one thousand qubits
and obtain the first  error threshold estimates for several toy models of coherent noise.
Numerical results are reported for storage of logical states subject to $Z$-rotation errors 
and for logical state preparation with general $SU(2)$ errors. 
We observe that for large code distances the effective logical-level noise
is well-approximated by random Pauli errors even though the physical-level noise
is coherent. 
Our algorithm works by mapping the surface code to a system of Majorana fermions. 
\end{abstract}

\maketitle

\section{Introduction}
\label{sec:intro}

Recent years have witnessed major progress  towards the 
demonstration of quantum error correction and reliable logical qubits~\cite{Barends2014,Kelly2015,corcoles2014,ofek2016demonstrating,takita2016demonstration}.
Topological  quantum codes such as the surface code~\cite{kitaev2003fault,bravyi1998quantum}
are  among the most attractive candidates for 
an experimental realization, as they can 
be implemented on a two-dimensional  grid of qubits
with local parity check operators.

It is believed  that such codes can tolerate a high level of
noise~\cite{Dennis2001,Raussendorf2007,Fowler2009}  which is
comparable to what can be achieved in the latest experiments~\cite{takita2016demonstration}.
 The general confidence in the noise-resilience of topological codes primarily 
rests on considerations of Pauli noise -- a simplified noise model where errors are Pauli operators
$X,Y,Z$ drawn at random from some distribution. An example is 
the case where each qubit~$j$ experiences noise described by the channel
\begin{align}
\mathcal{N}_j(\rho)&=(1-\epsilon)\rho+\epsilon_x X\rho X+\epsilon_y Y\rho Y+\epsilon_z Z\rho Z\ \label{eq:incoherentnoise}
\end{align}
with suitable probabilities $\epsilon_x,\epsilon_y,\epsilon_z$. 
This kind of noise can be fully described by  the stabilizer formalism~\cite{gottesman}. In pioneering work, Dennis et al.~\cite{Dennis2001}   exploited this algebraic structure to establish the first analytical threshold estimates,
see also~\cite{fowler2012proof}.
The effect of Pauli noise also is efficiently simulable thanks to the Gottesman-Knill theorem, providing numerical evidence for
high error thresholds  of topological codes~\cite{wang2010quantum}.
The efficient simulability property has recently been  extended beyond Pauli noise to  random Cliffords and 
Pauli-type  projectors~\cite{gutierrezclifford}.

While such algebraically defined noise models are attractive from  a theoretical viewpoint, they often do not correspond to noise encountered in real-world setups. They are -- in a sense -- not quantum enough: they model probablistic processes where errors act randomly on subsets of qubits.  Rather than being of such a probabilistic (or {\em incoherent}) nature, noise in a realistic device will often be coherent, i.e., unitary, and can involve small rotations acting everywhere. A typical situation where this arises is if e.g., frequencies of oscillator qubits are misaligned: this results in systematic unitary over- or 
under-rotations.
On a single-qubit level, this 
means that~\eqref{eq:incoherentnoise} should be replaced by noise of the form
\begin{align}
\mathcal{N}_j(\rho)&=U_j\rho U_j^\dagger \label{eq:coherentnoise}
\end{align}
with a suitable unitary operator $U_j\in SU(2)$. 
Since such  errors generally cannot be described within the stabilizer formalism, 
understanding their effect on a given quantum
 fault-tolerant scheme is a challenging problem. 

Prior theoretical work indicates  that  the difference between coherent and incoherent errors could be 
significant. In particular, it was observed~\cite{sanders,kueng,wallman,puzzuoli2014tractable,magesan2013modeling} that
coherent errors can lead to large differences between average-case and worst case fidelity 
measures suggesting that  a critical reassessment of  commonly used benchmarking measures is necessary.
This observation motivates the question of how much coherence is present in the effective
logical-level noise~\cite{rahn,fern2006generalized} experienced by encoded qubits.
Depending on whether or not the  logical noise is coherent one may choose different metrics
for quantifying performance of a given fault-tolerant scheme. 
Significant progress has been made towards understanding the structure of the logical noise
for concatenated codes~\cite{rahn,fern2006generalized,greenbaum2016modeling}.
However,  these studies are not directly applicable to large topological codes such as those
considered here.

Brute-force simulations of coherent noise in small codes
were presented in~\cite{gutierrez,TomitaSvore14,barnes,chamberland} for Steane codes and surface codes with up to~17 qubits. Simulating coherent errors by brute force  clearly requires time (and memory) exponential in the number of qubits~$n$. For the surface code, Darmawan and Poulin~\cite{darmawan} proposed an algorithm with a runtime exponential in~$n^{1/2}$ based on tensor networks, and simulated systems with up to 153~qubits. This algorithm can handle arbitrary noise (including e.g., amplitude damping). Unfortunately, its formidable  complexity prevents accurate estimation of error thresholds, e.g., for the systematic rotations considered here. In~\cite{suzuki}, threshold estimates for the 1D~repetition code were obtained. To our knowledge, there are no analogous  threshold estimates for topological codes
subject to coherent noise.

\begin{figure}[hb]
\includegraphics[height=3cm]{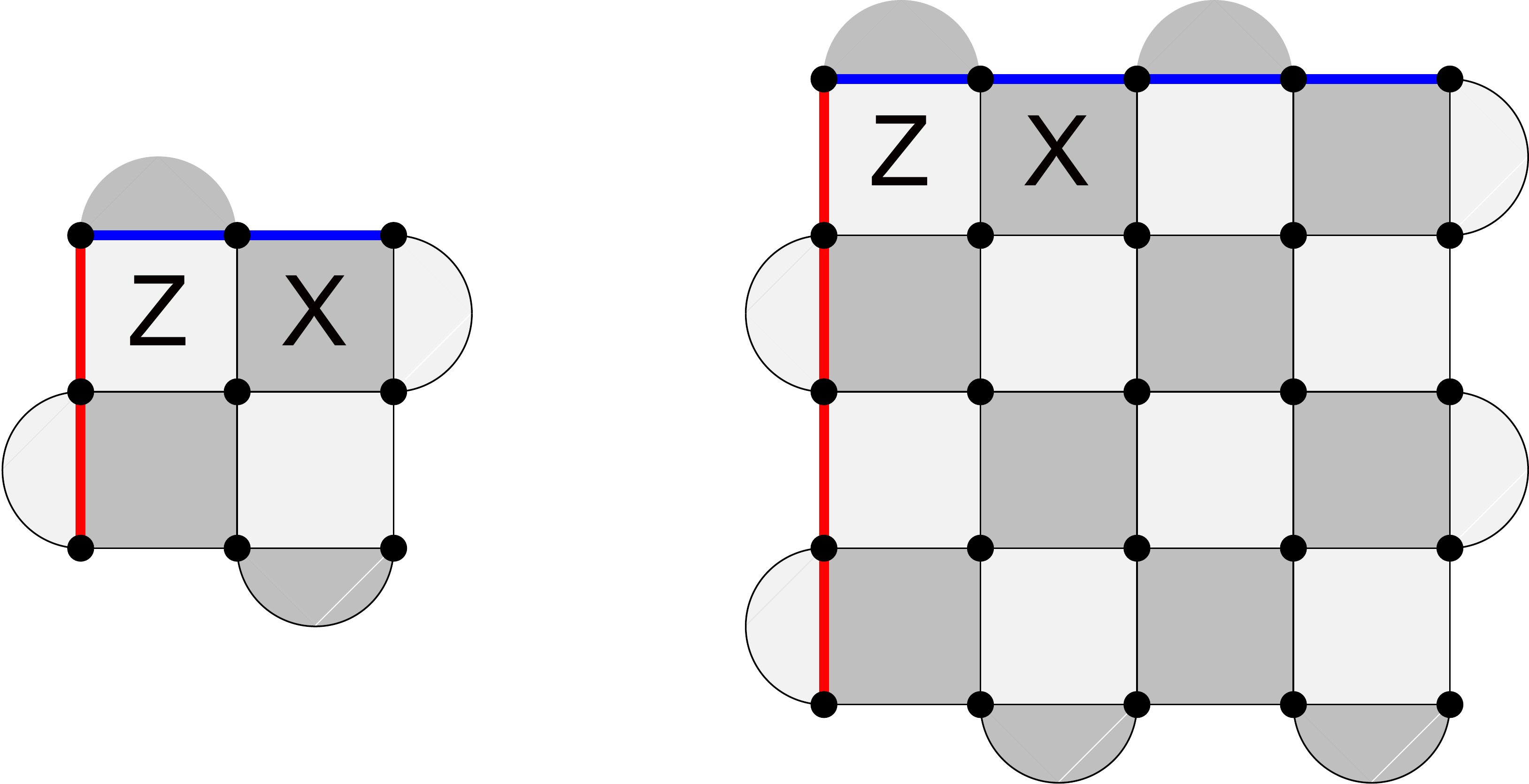}
\caption{Surface codes
with distance $d=3$ and $d=5$.
Qubits and stabilizers are located at sites and faces respectively.
A stabilizer $B_f$ located on a face $f$ 
applies $X$ (black faces) or $Z$ (white faces) to each qubit on the boundary of $f$. 
Logical Pauli operators $X_L$ (red) and $Z_L$ (blue) have support on
the left and the top boundary.
}
\label{fig:RSC}
\end{figure}

\paragraph*{Our setup.}
Here we show that  the effect of coherent errors in surface codes can be 
studied by means of polynomial-time algorithms. Specifically, we consider  coherent errors in the context of two 
central tasks associated with error correction, namely 
\begin{enumerate}[(A)]
\item fault-tolerant storage of quantum information.
\item fault-tolerant preparation of a logical basis state.
\end{enumerate}
We shall consider a particular version of the surface code proposed
in Refs.~\cite{wen2003quantum,Bombin2007optimal}.
A distance-$d$ surface code has one logical qubit and $n=d^2$ physical qubits
located at sites of a square lattice of size $d\times d$ with open boundary conditions.
The code has local stabilizers  $X^{\otimes 4}$, $X^{\otimes 2}$
or $Z^{\otimes 4}$, $Z^{\otimes 2}$ associated with
faces of the lattice as shown in Fig.~\ref{fig:RSC}. 
The stabilizer located on a face $f$ will be denoted $B_f$.
Logical Pauli operators $X_L$ and $Z_L$
acting on the encoded qubit  can be chosen as $X^{\otimes d}$ and $Z^{\otimes d}$ 
applied to the left and the top boundary of the lattice respectively.
The two-dimensional logical subspace is spanned by $n$-qubit states $\psi_L$
satisfying $B_f|\psi_L\rangle=|\psi_L\rangle$ for all $f$. 

To specify the problem~(A), consider a logical state $\psi_L$ initially encoded
by the surface code
and a coherent error $U=U_1\otimes \cdots \otimes U_n$
that applies some (unknown) unitary operator $U_j$ to each qubit $j$.
To diagnose and correct the error without disturbing the encoded state
we adopt the standard protocol
based on the syndrome measurement. It works by measuring the
eigenvalue (syndrome) $s_f=\pm 1$ of each stabilizer $B_f$ on the 
corrupted state $U|\psi_L\rangle$ and then applying a Pauli-type
correction operator $C_s$ depending on the measured syndrome $s=\{s_f\}_f$.
The correction $C_s$ is computed by a classical decoding algorithm
(for example, one may choose $C_s$ as a minimum-weight Pauli error consistent
with  $s$).
We note that the syndrome $s$ is a random variable with some probability distribution $p(s)$
since the error $U$ maps the initial logical state 
to a coherent superposition of states with different syndromes. 
In this paper we only consider noiseless syndrome measurements.
Accordingly, we assume that the correction $C_s$ always returns the system to the logical subspace
resulting in some final logical state $|\phi_s\rangle$.
For this problem we restrict  to $Z$-rotation errors, that is,
we assume that  $U_j=\exp(i \eta_j Z)$ for some (unknown) angles $\eta_j$.  
The restriction to $Z$-rotations is dictated by the  limitations of our simulation algorithm.
Thus we shall model a fault-tolerant storage by  the following process:
\begin{enumerate}[(i)]
\item prepare an  initial logical state $|\psi_L\rangle$
\item
apply a coherent error $\bigotimes_{j=1}^n \exp{(i\eta_j Z)}$ to $|\psi_L\rangle$
\item
 measure the eigenvalues of the stabilizers~$\{B_f\}_f$, resulting in a syndrome~$s=\{s_f\}_f$.
\item
apply a Pauli correction $C_s$ returning the system to the logical subspace
in some final state $|\phi_s\rangle$. 
\end{enumerate}
To assess how close the final state $|\phi_s\rangle$ and the initial state $|\psi_L\rangle$ are,
we seek a polynomial-time classical algorithm~$A$ which
takes as  input $|\psi_L\rangle$ and the rotation angles $\eta_1,\ldots,\eta_n$,  samples a syndrome~$s$ from the distribution 
$p(s)$ specified by the measurement~(iii), and outputs~$s$ as well as the associated final state~$|\phi_s\rangle$
(e.g. specified by its Bloch vector). By sampling sufficiently many syndromes,
one can learn how frequently and in which ways 
error correction may fail in the presence of coherent noise.

To specify the problem~(B), assume first that we have access to noise-free qubits and operations. In this case, the following standard protocol~\cite{Fowler2009} prepares the encoded stabilizer state~$|+_L\rangle$ (the $+1$ eigenstate of $X_L$):
\begin{enumerate}[(i)]
\item
prepare the initial product state $|+\rangle^{\otimes n}$.
\item measure the eigenvalues of the stabilizers~$\{B_f\}_f$, resulting in a syndrome~$s=\{s_f\}$.
\item apply a Pauli correction  $C_s$ returning the system to the logical subspace
in some final state $|\phi_s\rangle$.
\end{enumerate}
Using the fact the the initial  state is a $+1$ eigenvector of $X_L$ one
can easily check that $|\phi_s\rangle=|+_L\rangle$ for
all $s$, see~\cite{Fowler2009}.
How does this protocol fare in the presence of coherent noise? Let us consider a model where
the initial product state cannot be prepared 
with perfect accuracy, but rather is obtained from~$|+\rangle^{\otimes n}$ by
applying some unwanted unitary operators to every qubit. Thus~$(i)$ is replaced by
\begin{enumerate}[(i')]
\item
prepare an initial state $|\psi_1\otimes\psi_2\otimes\cdots \otimes \psi_n\rangle$, where $\psi_j$ 
are arbitrary single-qubit pure states. 
\end{enumerate}
In this case  the final  state~$|\phi_s\rangle$ may deviate from the target state~$|+_L\rangle$
resulting in a logical error.
To assess the performance of this protocol, we seek a polynomial time
algorithm $B$ which,  on input $\psi_1,\ldots,\psi_n\in \mathbb{C}^2$, 
outputs a random syndrome~$s$ sampled from the distribution~$p(s)$ specified by the measurement (ii)  together with the final logical state $|\phi_s\rangle$.

\paragraph*{Our results.}
We construct algorithms~A and B accomplishing the simulation tasks specified above. 
The runtime of  these algorithms scales as $O(n^2)$, where  we measure complexity in terms  of the number of additions, multiplications, and divisions on complex numbers that are required~\footnote{Strictly speaking, the simulation time scales as $O(n^2)+t(n)$, where $t(n)$
is the runtime of the decoding algorithm that computes the correction $C_s$. In our simulations the decoding time
was negligible compared with the time required to sample the syndrome and compute the final logical state.}.
Using these algorithms, we perform the first numerical study of large topological codes subject to coherent noise, performing simulations for  surface codes with up to $n=2401$~physical qubits,
 see Table~\ref{table:timing} for a timing analysis. 
 This shows that  efficient classical simulation of these fault-tolerance processes is possible,
and allows us to extract key characteristics of these codes in the limit of large system size.

We apply algorithm~A to study the effect of coherent noise on storage in the surface code.
We show that the syndrome probability distribution $p(s)$ is independent of the initial logical state $\psi_L$
whereas the final logical state has the form 
\begin{equation}
\label{theta_s_definition}
|\phi_s\rangle=\exp{(i\theta_s Z_L)} |\psi_L\rangle
\end{equation}
for some logical rotation angle $\theta_s\in [0,\pi)$ depending on the syndrome $s$.
We use the quantity
\begin{align}
\label{PLmem}
P^L &=2 \sum_s p(s)|\sin \theta_s|
\end{align}
as a measure of the logical error rate.  We will see that $P^L$
is the average diamond-norm distance between the conditional logical channel 
$\rho\mapsto e^{i\theta_s Z}\rho e^{-i\theta_s Z}$ and the identity channel.
For numerical simulations we consider translation-invariant coherent noise
of the form $(e^{i\theta Z})^{\otimes n}$, where $\theta\in[0,\pi)$ is the only noise parameter. 
The Pauli correction $C_s$ was computed using the standard
minimum weight matching decoder~\cite{Dennis2001,fowler2012towards} with 
constant weights independent of $\theta$. 
We are interested in the {\em error threshold}, that is, the maximum value $\theta_0$ such that
for any $\theta<\theta_0$ the logical error rate $P^L$ goes to zero in the limit $n\to \infty$.
We find the numerical estimate
\begin{align}
\label{threshold1}
0.08\pi\leq \theta_0\leq 0.1\pi.
\end{align}
Our numerical experiments confirm that, as expected, the quantity $P^L$ decays exponentially  in the code distance for values $\theta<\theta_0$ below the threshold. 
Surprisingly, the threshold estimate Eq.~(\ref{threshold1}) agrees very well
with the so-called Pauli twirl approximation~\cite{emerson2007symmetrized,silva2008scalable}
where coherent noise of the form~$\cN(\rho)=e^{i\theta Z}\rho e^{-i\theta Z}$ is replaced by 
dephasing noise ${\cal D}(\rho)=(1-\epsilon) \rho+ \epsilon Z\rho Z$,
with $\epsilon=\sin^2\theta$.
 For the latter the threshold error rate 
is around $\epsilon_0\approx 0.11$,
see Ref.~\cite{Dennis2001}.
Solving the equation
$\epsilon_0=\sin^2{(\theta_0)}$ for $\theta_0$
yields  $\theta_0 \approx 0.10\pi$, in agreement with Eq.~(\ref{threshold1}).
At the same time, we observe that the Pauli twirl approximation 
significantly underestimates $P^L$ in the sub-threshold regime, confirming that coherence
of noise may have a profound effect on a given fault-tolerant scheme, as was previously observed in~\cite{greenbaum2016modeling,darmawan}.

Algorithm~A allows us to investigate the probability distribution of logical rotation angle $\theta_s$
defined in Eq.~(\ref{theta_s_definition}).
We find that for large code sizes, this distribution concentrates around the two points $\{0,\pi/2\}$
which correspond to  the  logical Pauli-type  errors $\{I,Z_L\}$.
 To get a deeper insight into this phenomenon, 
we introduce and numerically study associated measures of ``incoherence''.
Our findings support the general conjecture that in the limit of large code distances, coherent physical noise gets converted into incoherent logical-level noise.

We apply Algorithm~B to study the effect of coherent noise on logical state preparation
in the case when the initial product state has the form 
\[
(\exp{(i\varphi X)} \exp{(i\theta Z)} |+\rangle)^{\otimes n}
\]
Here the angles
$\theta,\varphi\in [0,\pi)$  specify the action
of a coherent  error on the ideal initial state $|+\rangle$.
The ideal protocol corresponds to $\theta=0$. 
We define the logical error rate $P^L$ as the average trace-norm distance
between the final logical state $\phi_s$ and the target state $|+_L\rangle$,
see Section~\ref{subs:lspnum} for details. 
Our numerical results indicate that the error threshold can be described
by a single function $\theta_0(\varphi)$ such that  the logical error rate $P^L$ goes to zero
in the limit $n\to \infty$ for any $0\le \theta<\theta_0(\varphi)$
and $P^L$ is lower bounded by a positive constant for $\theta>\theta_0(\varphi)$.
We find the numerical estimate
\begin{equation}
\label{threshold2}
0.1\pi\leq \theta_0(\varphi)\leq 0.15\pi
\end{equation}
for all $\varphi$. This indicates that  the threshold funciton $\theta_0(\varphi)$ has a very mild (if any) dependence on 
$\varphi$.
We investigate the behavior of $P^L$ in more detail for $\varphi=0$
and obtain a  more refined estimate $0.13\pi\leq \theta_0(0)\leq 0.14\pi$.
The quantity $P^L$ is observed to decay exponentially in the code distance
in the sub-threshold regime. 

\begin{table}[h]
\begin{tabular}{r|c|c|c|c|c|}
Code distance & $\bf 9$ & $\bf 19$ & $\bf 29$ & $\bf 39$ & $\bf 49$  \\
\hline 
Qubits & $81$ & $361$ & $841$ & $1521$ & $2401$ \\
\hline
Runtime (A)& $0.001$ & $0.04$  & $0.2$  & $0.7$  & $1.7$   \\
\hline
Runtime (B) & $0.001$ & $0.01$ & $0.04$ &  $0.15$ & $0.4$    \\
\hline
\end{tabular}
\caption{Runtime in seconds for a C++ implementation of
algorithms~A and B. Timing analysis was performed on a laptop with a    2.6GHz Intel~i5 Dual Core CPU.
}
\label{table:timing}
\end{table}

\paragraph*{Outline.}
The remainder of this paper is structured as follows. 
Section~\ref{sec:methods} provides a high-level overview of our simulation algorithms.
In Section~\ref{sec:maj}, we describe a representation of the surface code
in terms of Majorana fermions.  In Sections~\ref{sec:prep},\ref{sec:store}, we give the classical simulation algorithms~A and~B and analyze their complexity. In Section~\ref{sec:numerics} we discuss our numerical results. We conclude in Section~\ref{sec:conclusions}.
Appendix~\ref{app:rotation} contains a proof of a technical lemma.
We provide some background on Majorana fermions and fermionic linear optics in 
Appendix~\ref{sec:gaussian}. 

\section{Methods}
\label{sec:methods}

Our main tool is a fermionic representation of the surface code proposed
by Kitaev~\cite{kitaev2006anyons} and Wen~\cite{wen2003quantum}.
It works by encoding each qubit of the surface code into four Majorana fermions
in a way that simplifies the structure of the surface code stabilizers. The Kitaev-Wen representation
has previously been  used by Terhal et al.~\cite{terhal2012majorana}  to design fermionic Hamiltonians with topologically
ordered ground states.
Here we show that this representation is also well-suited for  the design of 
efficient simulation algorithms. 
The fermionic version of the surface code will be described in terms of
Majorana operators $c_1,\ldots,c_{4n}$ that obey the standard commutation
rules $c_p^\dag=c_p$, $c_p^2=I$, and $c_pc_q=-c_qc_p$ for $p\ne q$. 
We will show that the 
error correction protocols considered in this paper can be 
decomposed into a sequence of  $O(n)$ elementary gates from a gate set 
known as a  {\em fermionic linear optics} (FLO), see 
Refs.~\cite{knill2001fermionic,bravyi2004lagrangian}.
 It includes the following operations:
\begin{enumerate}
\item Initialize a pair of Majorana modes $p,q$ in a basis state $|0\rangle$ satisfying $ic_pc_q |0\rangle=|0\rangle$.
\item Apply the unitary operator $U=\exp{(\gamma c_p c_q)}$. Here $\gamma\in [0,\pi)$ is a rotation angle.
\item Apply the projector $\Lambda=(I+ic_pc_q)/2$. Compute the norm of the resulting state.
\end{enumerate}
It is well-known that quantum circuits   composed of FLO gates
can be efficiently simulated classically~\cite{knill2001fermionic,terhal2002classical,bravyi2004lagrangian,bravyi2011classical}.
The simulation runtime scales as $O(n)$ for gates of type~(1,2) and as $O(n^2)$ for gates of type~(3). 
For completeness, we describe the requisite simulation algorithms in Appendix~\ref{sec:gaussian}. 
By exploiting the geometrically local structure of the surface  code we shall be able to reduce
the number of modes such that at any given time step the simulator only needs to keep track 
of $O(n^{1/2})$ modes.  Accordingly, each FLO gate can be simulated in time at most $O(n)$. Since the total number of gates is~$O(n)$, the total simulation time scales as $O(n^2)$.

\section{From qubits to Majorana fermions}
\label{sec:maj}

A single Majorana mode $p$ is described by a hermitian
operator $c_p$ satisfying $c_p^2=I$. Operators
$c_p$ associated with different modes anti-commute, see  
Appendix~\ref{sec:gaussian}
for formal definitions. 
A system of four Majorana modes $c_1,c_2,c_3,c_4$ can be used
to encode a qubit using a stabilizer code with a single stabilizer 
\begin{equation}
\label{C4stab}
S=-c_1c_2c_3c_4
\end{equation}
and logical Pauli operators 
\begin{equation}
\label{C4logical}
\overline{X}=ic_1c_2=ic_3c_4 S\quad \mbox{and} \quad \overline{Z}=ic_2c_3=ic_1c_4S.
\end{equation}
We shall refer to this encoding as a {\em $C4$-code}.

\begin{figure}[ht]
\includegraphics[height=3.5cm]{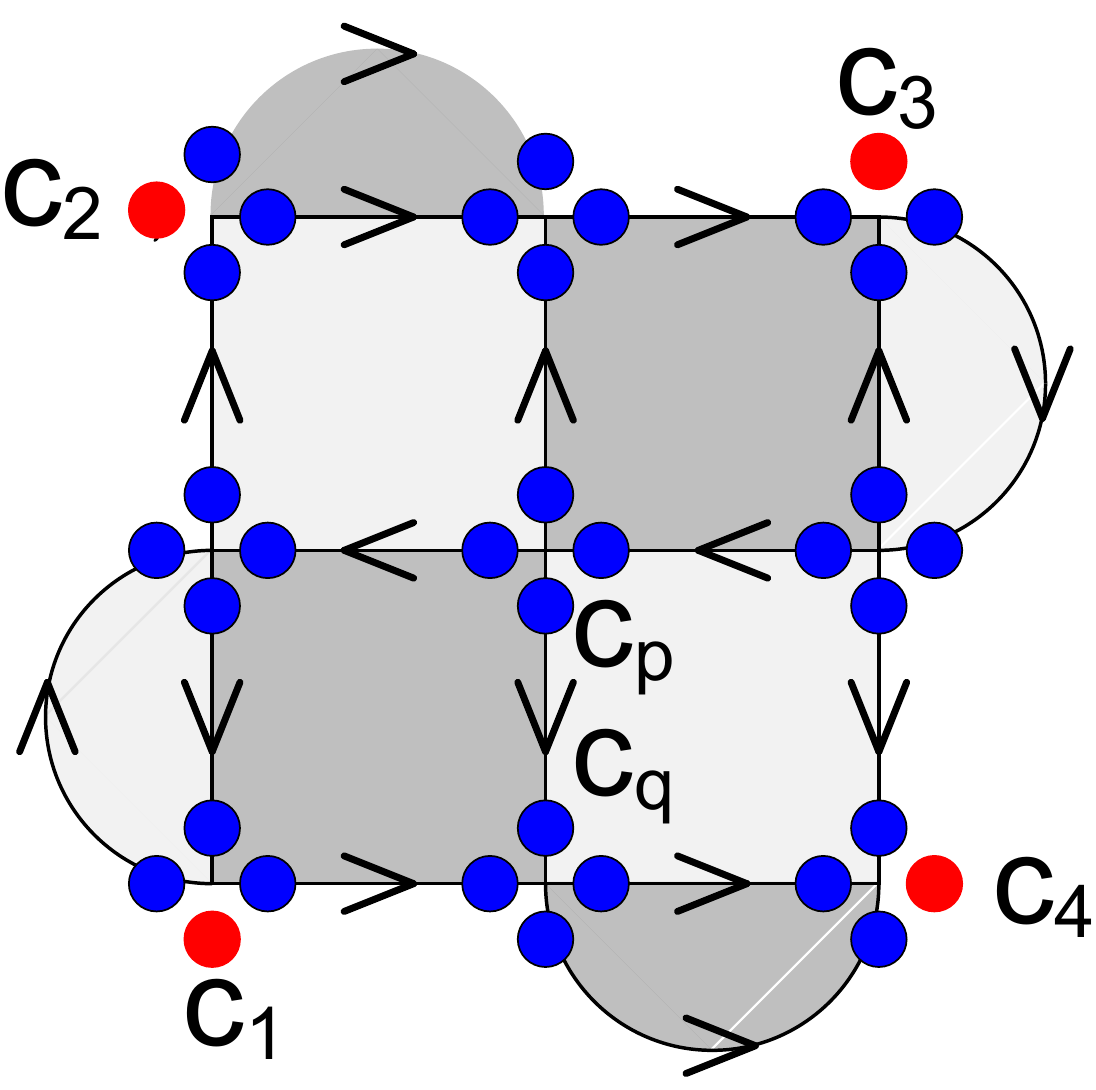}
\caption{Solid circles
represent paired (blue) and unpaired (red) Majorana modes. 
An edge oriented from $c_p$ to $c_q$
defines a link operator $L_e=ic_pc_q$.
The pattern extends to larger codes in a translation invariant fashion.
}
\label{fig:arrows}
\end{figure}

Consider a surface code with $n$ qubits on a square $d\times d$ lattice,
where $n=d^2$. It can  described by a planar graph
$G=(V,E,F)$ with a set of $n$ vertices $V$, a set of $2n-2$ edges $E$,
and a set of $n-1$ faces $F$. Qubits are located at vertices  $u\in V$
and stabilizers $B_f$ are located at faces $f\in F$  of $G$. 
Consider a system of $4n$ Majorana modes $c_1,\ldots,c_{4n}$
distributed over edges and vertices of $G$ as shown on Fig.~\ref{fig:arrows}.
There are exactly two {\em paired} modes located near the endpoints of every edge  $e\in E$
and four {\em unpaired} modes $c_1,c_2,c_3,c_4$ located near the corners of the lattice
as shown on Fig.~\ref{fig:arrows}. 
The paired modes are labeled as $c_5,c_6,\ldots,c_{4n}$ in an arbitrary order. 

Let us orient the edges of $G$ as shown on Fig.~\ref{fig:arrows}.
Suppose $e\in E$ is an edge 
connecting some pair of  modes $c_p$,  $c_q$
such that $c_p$ is the tail of $e$ and $c_q$ is the head of $e$, see Fig.~\ref{fig:arrows}.
Define the {\em link operator}
\begin{equation}
\label{link}
L_e=ic_p c_q.
\end{equation}
Note that $L_e$ is hermitian and all link operators pairwise
commute. Furthermore, each~$L_e$ commutes with the  unpaired modes $c_1,c_2,c_3,c_4$.

\begin{figure}[ht]
\includegraphics[height=3.5cm]{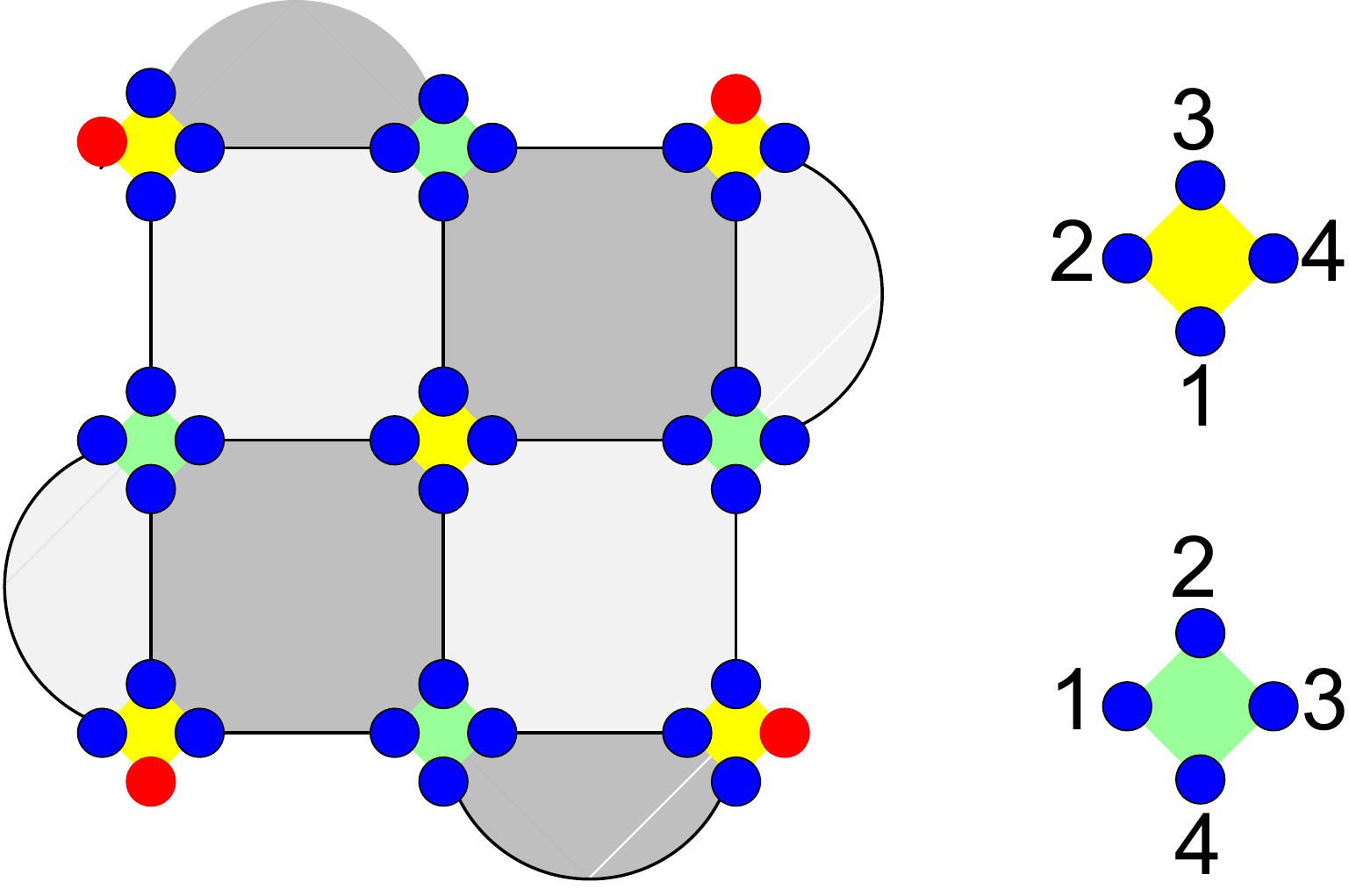}
\caption{Four-mode clusters $\Gamma_u$ located at vertices
 encode qubits of the surface code.
Modes in each cluster are ordered as shown on the right. 
Each vertex $u$ has a stabilizer $S_u=-c_{u,1}c_{u,2} c_{u,3} c_{u,4}$,
where $c_{u,j}$ is the $j$-th mode of $\Gamma_u$.
Logical Pauli operators are $\overline{X}_u=ic_{u,1}c_{u,2}=ic_{u,3}c_{u,4}S_u$
and $\overline{Z}_u=ic_{u,2}c_{u,3} = ic_{u,1} c_{u,4} S_u$,
see also Fig.~\ref{fig:arrows_all}.}
\label{fig:vertices}
\end{figure}

By construction, a small neighborhood of each vertex $u\in V$
contains a cluster of four modes, see Fig.~\ref{fig:vertices}.
 We shall denote this cluster $\Gamma_u$.
Define a {\em vertex stabilizer}
\begin{equation}
\label{Su}
S_u=-\prod_{p\in \Gamma_u} c_p,
\end{equation}
where a particular product order is chosen for each vertex
as shown on Fig.~\ref{fig:vertices}.
Since $|\Gamma_u|=4$ for all $u$ and the subsets  $\Gamma_u$ are pairwise disjoint,
 vertex stabilizers are hermitian and pairwise commuting.
We shall consider $S_1,\ldots,S_n$ as stabilizers for $n$ independent copies  of the $C4$-code defined
in Eqs.~(\ref{C4stab},\ref{C4logical})
such that each qubit of the surface code is encoded into its own $C4$-code. 
Let $\overline{X}_u$, $\overline{Z}_u$, and $\overline{Y}_u=i \overline{X}_u\overline{Z}_u$
be the logical Pauli operators
for the qubit located at a vertex $u$, see Eq.~(\ref{C4logical}).
By definition, each of these logical operators has the form $ i c_p c_q$
for some pair of Majorana modes $p,q\in \Gamma_u$, see Fig.~\ref{fig:vertices}.
The logical operators $\overline{X}_u$, $\overline{Z}_u$ are indicated by small arrows
on Fig.~\ref{fig:arrows_all}.

Let $P$ be a Pauli operator acting on the surface code qubits.
We shall say that a Majorana operator $\overline{P}$ is a $C4$-encoding
of $P$ if $\overline{P}$ can be obtained from $P$ by replacing
each single-qubit Pauli operator $X_u,Y_u,Z_u$ by its logical counterpart
$\overline{X}_u$, $\overline{Y}_u$,  $\overline{Z}_u$ and, possibly, multiplying by
the stabilizer $S_u$. 
Given a single-qubit state $\psi$, one can define several encoded versions 
of $\psi$ using the surface code, the $C4$-code, and the surface code
concatenated with $n$-copies of the $C4$-code.
We shall denote these encoded states $\psi_L$, $\overline{\psi}$, and $\overline{\psi}_L$
respectively. These notations are summarized in Table~\ref{table:encoding}.
\begin{center}
\begin{table}[h]
\begin{tabular}{c|c|c}
    & Hilbert space & Encoding \\
\hline
$\psi_L$ & $n$ qubits & surface code \\
\hline
$\overline{\psi}$ & four Majorana modes & $C4$-code \\
\hline
$\overline{\psi}_L$ & $4n$ Majorana modes & 
\parbox[t]{5cm}{encode each qubit of $\psi_L$ into  the $C4$-code} \\
\end{tabular}
\caption{Encoded versions of a single-qubit state $\psi$.}
\label{table:encoding}
\end{table}
\end{center}

\begin{figure}[htb]
\includegraphics[height=2.5cm]{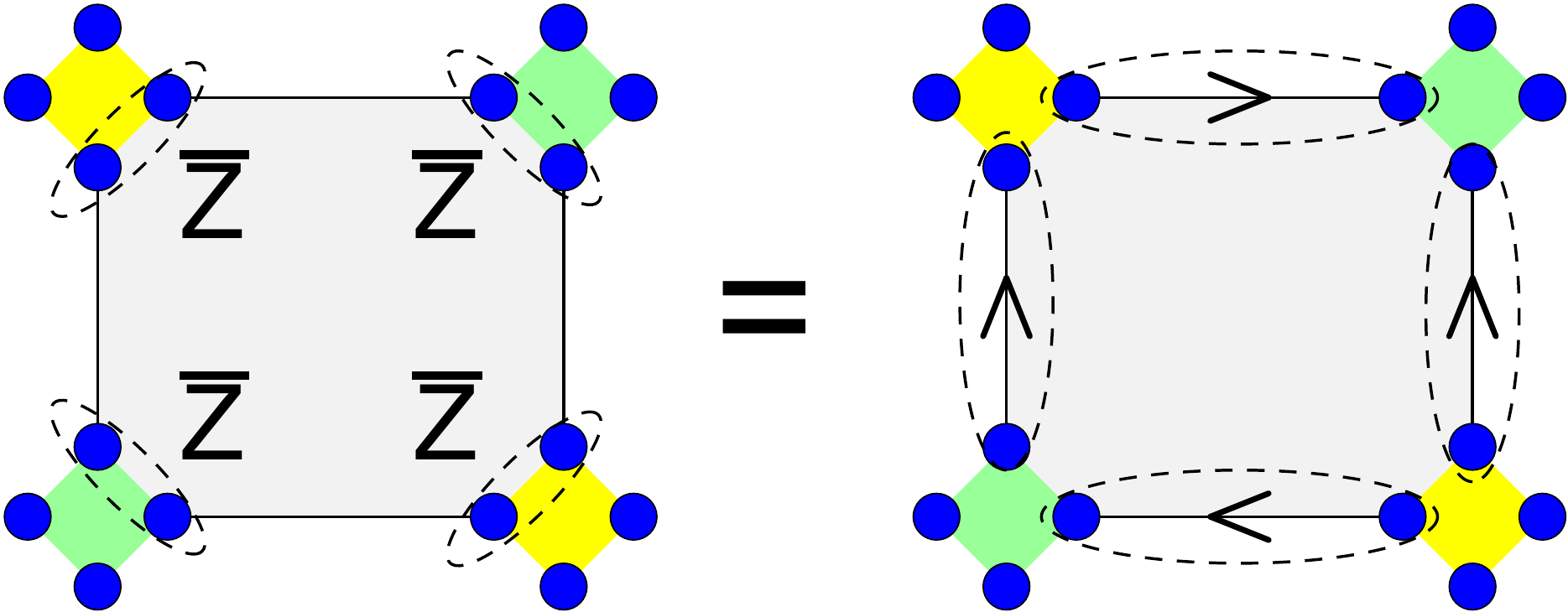}
\caption{{\em Left:} $C4$-encoding of  
a $Z$-type surface code stabilizer $B_f$. 
{\em Right:}  The same operator represented as a product of link operators $L_e$ over the boundary of $f$. 
}
\label{fig:plaq_c4}
\end{figure}

The desired  fermionic representation of the surface code is established in 
Lemmas~\ref{lemma:stab},\ref{lemma:logical},\ref{lemma:projection}
below.  Consider a face $f\in F$ and let $\partial f\subseteq E$ be the boundary of $f$.
\begin{lemma}
\label{lemma:stab}
Let $B_f$ be the surface code stabilizer located on a face $f$.
Then a $C4$-encoding of $B_f$ can be chosen as 
\begin{equation}
\label{Bbar}
\overline{B}_f = \prod_{e\in \partial f} L_e.
\end{equation}
\end{lemma}
We illustrate Eq.~(\ref{Bbar}) on Fig.~\ref{fig:plaq_c4}.
The lemma shows that measuring the surface code syndrome
can be reduced (after the $C4$-encoding) to measuring eigenvalues of 
pairwise commuting  link operators $L_e$. We shall see that under certain circumstances
such measurements can  be efficiently simulated classically.

\begin{figure}[htb]
\includegraphics[height=2cm]{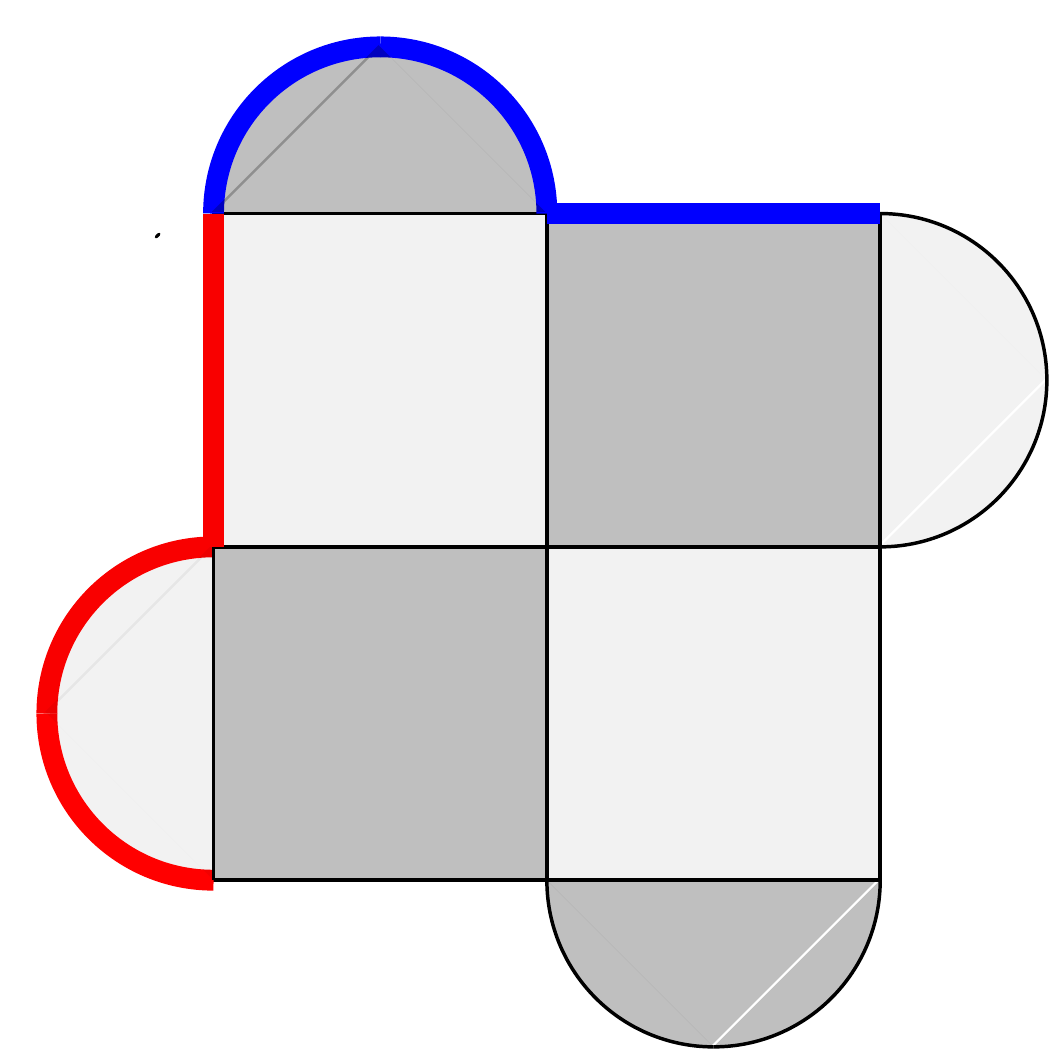}
\caption{The sets of edges LEFT (red) and TOP (blue). }
\label{fig:logicalXZ}
\end{figure}

\begin{proof}
Consider a face $f$ such that 
$B_f$ is a $Z$-stabilizer.
Then $\overline{B}_f = \prod_{u\in f} \overline{Z}_u$.
Consider a vertex $u\in f$ and the $C4$-code located at $u$.
The corresponding 
logical-$Z$ operator $\overline{Z}_u=i c_p c_q$ can be chosen such that both modes
$c_p$, $c_q$ are located on  the boundary of  $f$, see Fig.~\ref{fig:vertices}.
Thus $\overline{B}_f$ is proportional to the product of all 
modes located on the boundary of $f$. 
The same is true about the operator $\prod_{e\in \partial f} L_e$.
Thus
$\prod_{e\in \partial f} L_e=\pm  \overline{B}_f$.

\begin{figure}[htb]
\includegraphics[height=5cm]{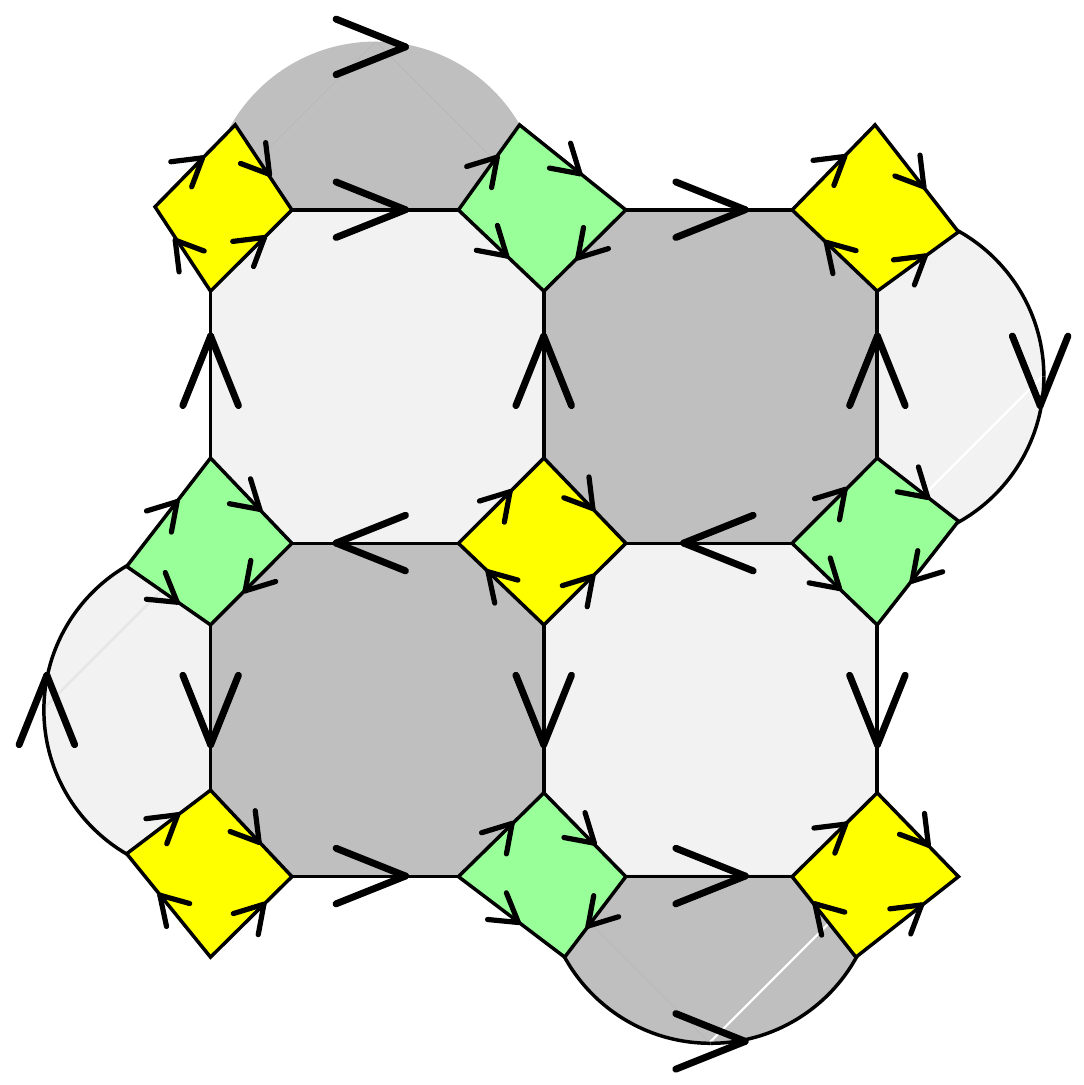}
\caption{Small arrows indicate the order of modes in the logical
operators $\overline{X}_u$ and $\overline{Z}_u$.
Each of these operators has the form $ic_pc_q$, where $p$ is the tail
and $q$ is the head of the respective arrow. The  orientation
is chosen such that  the boundary of each face has an odd number of arrows 
oriented clockwise. To avoid clutter, we do not show the Majorana modes. 
}
\label{fig:arrows_all}
\end{figure}

By construction, the boundary  $\partial f$ alternates between
link operators $L_e$ and logical-$Z$ operators $\overline{Z}_u$,
see Fig.~\ref{fig:plaq_c4} for an example.
For each link operator $L_e$ with $e\in \partial f$
define a quantity $\omega_f(L_e)=\pm 1$ such that 
$\omega_f(L_e)=-1$ iff $e$ is oriented clockwise with respect to $f$.
Likewise, for each logical-$Z$ operator $\overline{Z}_u=ic_pc_q$ lying on the boundary of $f$
define a quantity $\omega_f(\overline{Z}_u)=\pm 1$ such that 
$\omega_f(\overline{Z}_u)=-1$ iff an arrow  $c_p\to c_q$ is oriented clockwise with respect to $f$.
See Fig.~\ref{fig:arrows_all} for examples of such arrows.
A simple computation shows that 
$\prod_{e\in \partial f} L_e= -\omega_f \overline{B}_f$, where
\[
\omega_f=\prod_{e\in \partial f} \omega_f(L_e) \prod_{u\in f} \omega_f(\overline{Z}_u).
\]
Thus we need to check that $\omega_f=-1$ for each face $f\in F$. In other words,
the boundary of each face must have an odd number of arrows oriented clockwise.
 Direct inspection shows that this is indeed the case for the distance-$3$ code,
see Fig.~\ref{fig:arrows_all}.
By translation invariance, this also holds for all code distances. 
The same arguments apply to $X$-type stabilizers.
\end{proof}
We shall need an analogue of Lemma~\ref{lemma:stab} for logical operators of the surface code. 
\begin{lemma}
\label{lemma:logical}
Let $X_L$ and $Z_L$ be the logical operators of the surface code located
on the left and the top boundary. 
Then $C4$-encodings of $X_L$ and $Z_L$ can be chosen as 
\begin{equation}
\label{logicalXZ}
\overline{X}_L=ic_1c_2 \prod_{e\in \mathrm{LEFT}} L_e
\quad \mbox{and} \quad \overline{Z}_L=ic_2c_3 \prod_{e\in \mathrm{TOP}} L_e,
\end{equation}
where LEFT and TOP are the subsets of edges lying on the left
and the top boundaries of the lattice, see Fig.~\ref{fig:logicalXZ}.
\end{lemma}
\begin{proof}
Let us add a ``logical edge" connecting the modes $c_2$ and $c_3$ to the graph $G$,
see Fig.~\ref{fig:arrows}.
This creates an extra ``logical face" $f$ attached to the top boundary of the lattice.
The new edge carries a link operator $L_e=ic_2c_3$.
The same arguments as in the proof of Lemma~\ref{lemma:stab} show that 
$\overline{Z}_L=-\omega_f \prod_{e\in \partial f} L_e$, where  $\omega_f=\pm 1$
is the parity of the number of arrows lying on the boundary of $f$ and oriented clockwise
with respect to $f$. From Fig.~\ref{fig:arrows_all} one gets $\omega_f=-1$.
The same argument applies to $\overline{X}_L$.
\end{proof}

Suppose $\psi$ is a single-qubit state.
The following lemma allows one to switch between different encodings of $\psi$
defined in Table~\ref{table:encoding}
by measuring syndromes of the vertex stabilizers.
Informally, it asserts that a qubit  initially encoded into the four unpaired Majorana modes
$c_1,c_2,c_3,c_4$
at the corners of the lattice can be ``injected" into the logical subspace of the surface code
by tensoring in unentangled pairs of modes located on edges  and measuring
syndromes of the vertex stabilizers. 
\begin{lemma}
\label{lemma:projection}
Let $\phi_{\mathsf{link}}$ be the  state of $4n-4$ Majorana modes $c_5,c_6,\ldots,c_{4n}$
stabilized by all link operators, 
\begin{equation}
\label{phi_link}
|\phi_{\mathsf{link}}\rangle\langle \phi_{\mathsf{link}}|=\prod_{e\in E} \frac12 (I+L_e).
\end{equation}
Let $\psi$ be any single-qubit state. Then 
\begin{equation}
\label{projection}
|\overline{\psi}_L\rangle\sim \prod_{u\in V} \frac12(I+S_u)  |\overline{\psi}\rangle \otimes |\phi_{\mathsf{link}}\rangle.
\end{equation}
Here we used the notations from Table~\ref{table:encoding}.
\end{lemma}
The lemma will allow us to replace the initial logical state $\psi_L$ in the 
error correction protocol by a simpler state $\phi_{\mathsf{link}}$ at the cost of measuring
certain additional stabilizers. We shall see that the state $\phi_{\mathsf{link}}$
is a  fermionic Gaussian state, see Section~\ref{sec:gaussian} for details.
Furthermore, a state obtained from $\phi_{\mathsf{link}}$ by 
applying a coherent error (encoded by the $C4$-code) is also Gaussian.
These features will be instrumental for our simulation algorithm. 
\begin{proof}
Suppose first that $|\psi\rangle=|0\rangle$. Let us add  a pair of ``logical edges"
connecting modes $c_2,c_3$ and $c_1, c_4$ to the graph $G$.
This creates an extra pair of ``logical faces" attached to the top and the bottom boundaries.
The new edges carry link operators $ic_2c_3$ and $ic_1 c_4$. 
Lemmas~\ref{lemma:stab},\ref{lemma:logical} imply  that the state
$|\overline{\psi}\rangle \otimes |\phi_{\mathsf{link}}\rangle$ is stabilized by operators
$\overline{B}_f$ and $\overline{Z}_L$. Furthermore, since these operators
commute with all vertex stabilizers, the state on the right-hand side
of Eq.~(\ref{projection}) is stabilized by $\overline{B}_f$ and $\overline{Z}_L$.
Since it is also stabilized by all $S_u$, this state has the same set of stabilizers
as $|\overline{0}_L\rangle$, which proves Eq.~(\ref{projection}) for $|\psi\rangle=|0\rangle$.
Note that 
\[
\overline{X}_L |\overline{0}_L\rangle \otimes  |\phi_{\mathsf{link}}\rangle = |\overline{1}_L\rangle \otimes  |\phi_{\mathsf{link}}\rangle,
\]
see Lemma~\ref{lemma:logical}. Furthermore, 
$\overline{X}_L$ commutes with all vertex stabilizers. Thus applying $\overline{X}_L$
to both sides of Eq.~(\ref{projection}) with $|\psi\rangle=|0\rangle$ proves
Eq.~(\ref{projection}) for $|\psi\rangle=|1\rangle$. By linearity, it holds for all $\psi$.
\end{proof}

\section{Logical state preparation}
\label{sec:prep}

We shall first describe  the algorithm for simulating the logical state preparation 
because it is much simpler than the storage simulation.
For each syndrome $s=\{s_f\}_{f\in F}$ define a syndrome projector
\begin{equation}
\label{Ps}
\Pi_s = \prod_{f\in F} \frac12 (I+s_f B_f).
\end{equation}
It projects onto the subspace spanned by $n$-qubit states with the syndrome $s$.
Note that  $\sum_s \Pi_s=I$. 
Let $\Pi_0$ be the projector onto the logical subspace of the surface code.
Since the Pauli correction $C_s$ maps any state with a syndrome $s$
to the logical subspace, it must satisfy
\begin{equation}
\label{corr}
\Pi_s= C_s \Pi_0 C_s.
\end{equation}
Here and below we assume that $C_s^\dag=C_s$.

Suppose  that our initial state has the product form 
\[
|\psi\rangle =|\psi_1 \otimes \psi_2 \otimes \cdots \otimes \psi_n\rangle.
\]
Here $\psi_j$ are arbitrary single-qubit states. 
Our goal is to sample a syndrome $s$ from the probability distribution 
\begin{equation}
\label{prob1}
p(s)=\langle \psi |\Pi_s |\psi\rangle
\end{equation}
and compute  the final logical state conditioned on the syndrome.
The latter has the form 
\begin{equation}
\label{final1}
|\phi_s\rangle = \frac1{\sqrt{p(s)}} C_s \Pi_s |\psi\rangle
\end{equation}

Let us first discuss how to simulate the syndrome measurement. 
We shall encode each qubit into the $C4$-code as discussed in Section~\ref{sec:maj}.
Let $|\overline{\psi}_a\rangle$ be the encoded version of $|\psi_a\rangle$. 
Using the Euler angle decomposition one can write 
$|\psi_a\rangle = e^{-i \alpha X} e^{-i \beta Z}  e^{-i \gamma X} |0\rangle$ .
Replacing Pauli operators by their $C4$-encodings defined in Eq.~(\ref{C4logical}) gives
\[
|\overline{\psi}_a\rangle=\exp{(\alpha c_1 c_2)} \exp{(\beta c_2 c_3)} \exp{(\gamma c_1 c_2)} |\overline{0}\rangle.
\]
The state $|\overline{0}\rangle$ is stabilized by $\overline{Z}=ic_2c_3$ and $\overline{Z}S=ic_1c_4$,
see Eq.~(\ref{C4logical}). Thus the states $|\overline{\psi}_a\rangle$  
can be prepared using only FLO gates. Applying this independently to each qubit
gives a sequence of $O(n)$ FLO gates that 
prepares  the state
\[
|\overline{\psi}\rangle = |\overline{\psi}_1 \otimes \cdots \otimes \overline{\psi}_n\rangle.
\]

Clearly,  measuring syndromes of stabilizers $B_f$
on the state $|\psi\rangle$ is  equivalent to measuring syndromes of  the
encoded stabilizers $\overline{B}_f$ on the state $|\overline{\psi}\rangle$. 
By Lemma~\ref{lemma:stab},  the latter can be reduced to measuring
syndromes $m_e=\pm 1$ of the link operators $L_e$
and then classically computing face syndromes $s_f=\prod_{e\in \partial f} m_e$. 
Since each link operator has a form $L_e=ic_pc_q$,
measuring the eigenvalue of $L_e$ is a FLO gate.
Thus 
we have realized the full syndrome measurement using $O(n)$ FLO gates. 

It remains to compute the final logical state $\phi_s$.
Let $\vec{b}_s=(b^x_s,b^y_s,b^z_s)$ be the Bloch vector of $\phi_s$ such that 
\[
b^x_s=\langle \phi_s| X_L|\phi_s\rangle,
\quad b^y_s=\langle \phi_s| Y_L|\phi_s\rangle,\quad 
b^z_s=\langle \phi_s| Z_L|\phi_s\rangle.
\]
Below we focus on computing $b^x_s$. Define $\lambda_s=\pm 1$ such that 
$C_s X_L=\lambda_s X_L C_s$. 
 Then 
\begin{equation}
\label{bX}
b^x_s= \lambda_s  \frac{\langle \psi| \Pi_s X_L |\psi\rangle}{\langle \psi|\Pi_s|\psi\rangle}=
\lambda_s  \frac{\langle \overline{\psi} | \overline{\Pi}_s \overline{X}_L |\overline{\psi}\rangle}
{\langle \overline{\psi}|\overline{\Pi}_s|\overline{\psi}\rangle}
\end{equation}
Here in the first equality we noted that $X_L$ commutes with $\Pi_s$. In the second equality we  encoded
every qubit using the $C4$-code.
Let $m\in \{+1,-1\}^E$ be the combined syndrome of link operators measured in the
first part of the algorithm such that $m_e$
is the measured eigenvalue of $L_e$. Define the corresponding syndrome projector
\[
\Pi^{\mathsf{link}}_m = \prod_{e\in E} \frac12 (I  + m_e L_e).
\] 
We claim that 
\begin{equation}
\label{bX2}
b^x_s=
\lambda_s \prod_{e\in \mathrm{LEFT}} m_e\cdot 
\frac{\langle \overline{\psi} | {\Pi}_m^{\mathsf{link}}  ic_1 c_2  |\overline{\psi}\rangle}
{\langle \overline{\psi}|{\Pi}_m^{\mathsf{link}} |\overline{\psi}\rangle}.
\end{equation}
Here LEFT is the subset of edges lying on the left boundary of the lattice,
see Fig.~\ref{fig:logicalXZ}.
The ratio in Eq.~(\ref{bX2}) can be computed 
by taking the normalized state ${\Pi}_m^{\mathsf{link}} |\overline{\psi}\rangle$
obtained  after
measuring the link syndromes (the first part of the algorithm)
and measuring the eigenvalue of $ic_1c_2$.
The latter requires a  single FLO gate.
This gives the desired value of $b^x_s$.
Likewise, measuring the eigenvalues of $ic_2c_3$
and $-ic_1c_3$ on the final state
${\Pi}_m^{\mathsf{link}} |\overline{\psi}\rangle$
gives the remaining components of the Bloch vector
$b^z_s$ and $b^y_s$ respectively.

To conclude, FLO gates enable simulation of  the syndrome measurement
and computation of the final logical Bloch vector conditioned on the measured  syndrome. 
The resulting FLO circuit can be simulated classically in time $O(n^3)$ since it includes
$O(n)$ projection gates $(1/2)(I+m_e L_e)$.
 The simulation runtime can be significantly improved
using the following simple observations. 
First, once a link operator $L_e=ic_pc_q$ has been
measured, the modes $c_p$, $c_q$ are completely disentangled from the rest of the system.
Such disentangled modes can be removed from the simulator reducing the total number
of modes and the computational cost of subsequent steps.
Second, 
we can  exploit the fact that the  initial state $\overline{\psi}$ has a product form.
In particular, a four-mode cluster $\Gamma_u$ that supports the state
$\overline{\psi}_u$ only needs to be loaded into the simulator at a time step
when some mode $p\in \Gamma_u$ participates in the measurement of a link operator.
Thus at any given time step the simulator only needs to keep track of ``active" 
clusters $\Gamma_u$ such that at least one mode $p\in \Gamma_u$ has been measured
and at least one mode $q\in \Gamma_u$ has not been measured. 
One can easily choose the order of measurements such that the number of active clusters
is $O(n^{1/2})$ at any time step (for example, one can measure link operators column by column).
Accordingly, the cost of simulating a single FLO gate is at most $O(n)$.
Since the number of gates is $O(n)$, the total simulation cost is $O(n^2)$.

It remains to prove Eq.~(\ref{bX2}).
Let $\delta$  be a map from  link syndromes to the corresponding face syndromes, that is,
$s=\delta(m)$ iff $s_f=\prod_{e\in \partial f} m_e$ for all $f\in F$. 
By Lemma~\ref{lemma:stab}, 
\begin{equation}
\label{aaa1}
\overline{\Pi}_s=\sum_{m \, : \, \delta(m)=s} \Pi^{\mathsf{link}}_s.
\end{equation}
Below we prove the following
\begin{prop}
\label{prop:link}
Suppose $m$ and $m'$ are link syndromes such that $\delta(m)=\delta(m')$.
Then there exists a subset of vertices $W\subseteq V$ such that 
\[
\Pi_{m'}^{\mathsf{link}}= T \Pi_{m}^{\mathsf{link}}T, \quad \mbox{where} \quad T=\prod_{u\in W} S_u.
\]
\end{prop}
We postpone the proof until the end of the section. 
Let  $m$ and $m'$ be any link syndromes with $\delta(m)=\delta(m')$. Then
the proposition implies that
\begin{equation}
\label{aaa2}
\langle \overline{\psi}| {\Pi}_{m'}^{\mathsf{link}} |\overline{\psi}\rangle
=\langle \overline{\psi}| T {\Pi}_{m}^{\mathsf{link}}T  |\overline{\psi}\rangle
=\langle \overline{\psi}|{\Pi}_{m}^{\mathsf{link}} |\overline{\psi}\rangle
\end{equation} 
since $S_u\overline{\psi}=\overline{\psi}$  for all $u$. 
Likewise, 
\begin{equation}
\label{aaa3}
\langle \overline{\psi}| {\Pi}_{m'}^{\mathsf{link}} \overline{X}_L |\overline{\psi}\rangle
=\langle \overline{\psi}|{\Pi}_{m}^{\mathsf{link}}\overline{X}_L |\overline{\psi}\rangle
\end{equation}
since $ \overline{X}_L$ commutes with $T$. 
From Eqs.~(\ref{bX},\ref{aaa1},\ref{aaa2},\ref{aaa3}) one gets
\begin{equation}
\label{bX1}
b^x_s=
\lambda_s  \frac{\langle \overline{\psi} | {\Pi}_m^{\mathsf{link}} \overline{X}_L  |\overline{\psi}\rangle}
{\langle \overline{\psi}|{\Pi}_m^{\mathsf{link}} |\overline{\psi}\rangle}
\end{equation}
for any link syndrome $m$ such that $\delta(m)=s$.
In particular, one can choose $m$ as the link syndrome measured in the first part of the
algorithm.
Substituting $\overline{X}_L$ from 
Lemma~\ref{lemma:logical}  gives
the desired result Eq.~(\ref{bX2}).
It remains to prove Proposition~\ref{prop:link}.
\begin{proof}
By linearity, we can assume that $m$ is the trivial syndrome, that is, 
$m_e=1$ for all $e\in E$. Then 
$\delta(m')=\delta(m)$ iff
$m'$ is a flat connection on the surface code lattice,
that is, $m'$ is a $1$-chain with $\ZZ_2$ coefficients such that 
the $\ZZ_2$-valued magnetic flux through every face is $+1$.
Since the lattice is topologically trivial, any flat connection 
is a co-boundary of some $0$-chain $W\subseteq V$. 
Since a vertex stabilizer $S_u$ flips the link syndromes on all edges
incident to $u$, the condition that $m'$ is a co-boundary of $W$
is equivalent to
 $\Pi_{m'}^{\mathsf{link}}= T \Pi_{m}^{\mathsf{link}}T$
for $T=\prod_{u\in W} S_u$.
\end{proof}

\section{Storage of a logical state}
\label{sec:store}

Assume now that our initial state  has the form 
\[
|\psi\rangle = U |\psi_L\rangle, \qquad U=e^{i\eta_1 Z}\otimes e^{i\eta_2 Z} \otimes \cdots \otimes e^{i\eta_n Z}
\]
where $\psi_L$ is some  (unknown) logical state of the surface code
and $\eta_1,\ldots,\eta_n$ are arbitrary  rotation angles.
The unitary  $U$ describes a coherent error
applied to each qubit before the syndrome measurement.
Our goal is to sample a syndrome $s$ from the probability distribution 
\begin{equation}
\label{prob2}
p(s)=\langle \psi |\Pi_s |\psi\rangle
\end{equation}
and compute  the final logical state conditioned on the syndrome,
\begin{equation}
\label{final2}
|\phi_s\rangle = \frac1{\sqrt{p(s)}} C_s \Pi_s |\psi\rangle\ .
\end{equation}
Clearly, since $\psi$ contains only $Z$-type errors, the observed
syndrome of $Z$-stabilizers is always trivial.
Accordingly, we shall assume that the  correction $C_s$
is a $Z$-type Pauli. 
We shall need the following fact.
\begin{lemma}
\label{lemma:rotation}
The probability $p(s)$  does not depend on the initial logical state $\psi_L$.
The map $\psi_L\to \phi_s$  is a logical  $Z$-rotation by some angle
$\theta_s\in [0,\pi)$, that is,
\begin{equation}
\label{syn2}
|\phi_s\rangle =\exp{\left[ i \theta_s Z_L \right]}  |\psi_L\rangle.
\end{equation}
\end{lemma}
We shall refer to $\theta_s$ as a {\em logical rotation angle}.
Here and below all states are defined modulo an overall phase factor. 
We defer the proof of the lemma to Appendix~\ref{app:rotation}.

\subsection{Simulating the syndrome measurement}
\label{subs:syndrome}

\begin{figure}[hb]
\includegraphics[height=2cm]{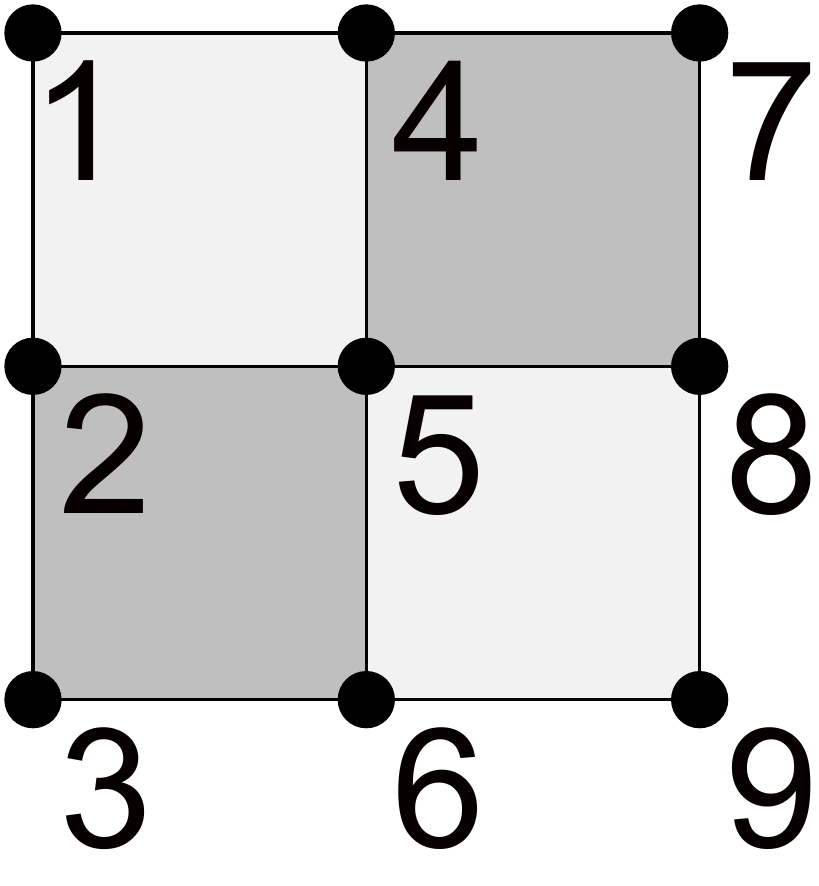}
\caption{Ordering of qubits. 
}
\label{fig:order}
\end{figure}

Let us first discuss how to sample $s$ from $p(s)$. 
By Lemma~\ref{lemma:rotation}, $p(s)$ does not depend on $\psi_L$, so  below we 
set $|\psi_L\rangle =|+_L\rangle$.
Since only syndromes of $X$-stabilizers may be non-trivial, 
it suffices to measure eigenvalues $m_u=\pm 1$ of single-qubit Pauli operators
$X_u$  and then classically compute the face syndrome $s_f=\prod_{u\in f} m_u$
for  each face $f$ that supports an $X$-stabilizer.
Let $m\in \{+1,-1\}^V$ be the combined $X$-measurement outcome and
$p^x(m)$ be the probability of an outcome $m$. 
We have
\begin{equation}
\label{memory1}
p^x(m)=\langle \psi |\Pi^x_m|\psi\rangle, \qquad \Pi^x_m \equiv \prod_{u\in V}
\frac12 (I+m_u X_u).
\end{equation}
Let us order qubits column by column as shown on Fig.~\ref{fig:order}.
Let $p^x_t(m_1,\ldots,m_t)$ be the marginal distribution of $p^x(m)$ describing the first $t$
qubits and let 
\begin{equation}
\label{conditional}
p^x_t(m_t|m_1,\ldots,m_{t-1}) = \frac{p^x_t(m_1,\ldots,m_t)}{p^x_{t-1}(m_1,\ldots,m_{t-1})}
\end{equation}
be the conditional distribution of $m_t$  given $m_1,\ldots,m_{t-1}$. 

Let us show how to sample $m_t$ from the distribution Eq.~(\ref{conditional})
using FLO gates. 
Partition the set of all qubits as $[n]=AB$ where 
$A=\{1,\ldots,t\}$ and $B=[n]\setminus A$.
We shall write 
\[
m_A=(m_1,\ldots,m_t), \quad U_A=\prod_{u\in A} e^{i\eta_u Z_u},
\]
and
\[
\Pi^x_A=\prod_{u\in A} \frac12(I+ m_u X_u).
\]
Then $p^x_t(m_A)=\langle \psi_L| U_A^\dag \Pi^x_A U_A |\psi_L\rangle$
Encoding each qubit into the $C4$-code as discussed in Section~\ref{sec:maj} gives
\begin{equation}
\label{memory3}
p^x_t(m_A)=\langle \overline{\psi}_L| \overline{U}_A^\dag \overline{\Pi}^x_A \overline{U}_A |\overline{\psi}_L\rangle.
\end{equation}
The Majorana representation of the logical state $\overline{\psi}_L$  defined in
Lemma~\ref{lemma:projection}  gives
\begin{equation}
\label{memory4}
|\overline{\psi}_L\rangle =\gamma^{1/2} \prod_{u\in V} \frac12(I+S_u) |\overline{\psi}_L\rangle \otimes |\phi_{\mathsf{link}}\rangle,
\end{equation}
where $\gamma$ is a normalizing coefficient depending only on $n$,
 $\phi_{\mathsf{link}}$ is a product state defined in Eq.~(\ref{phi_link}),
and $\overline{\psi}_L$ is the basis state of  modes $c_1,c_2,c_3,c_4$
stabilized by $ic_1c_2$ and $ic_3c_4$ (recall that we have chosen $|\psi_L\rangle=|+_L\rangle$).
The state $|\overline{\psi}_L\rangle \otimes |\phi_{\mathsf{link}}\rangle$ can be prepared using FLO gates
since it is stabilized by two-mode operators $L_e$, $ic_1c_2$ and $ic_3c_4$.
To simplify the notation, we shall absorb the pairs $ic_1c_2$ and $ic_3c_4$ into
$\phi_{\mathsf{link}}$. Accordingly, below we assume that 
$\phi_{\mathsf{link}}$ is a state of $4n$ modes defined as
\begin{equation}
\label{memory5}
|\phi_{\mathsf{link}}\rangle\langle \phi_{\mathsf{link}}|=\frac12(I+ic_1c_2) \frac12 (I+ic_3c_4) \prod_{e\in E} \frac12 (I+L_e).
\end{equation}
Plugging Eq.~(\ref{memory4}) into Eq.~(\ref{memory3}) gives
\begin{equation}
\label{memory6}
p^x_t(m_A)=\gamma \langle \phi_{\mathsf{link}} |\Omega \overline{U}_A^\dag \overline{\Pi}^x_A \overline{U}_A \Omega|
\phi_{\mathsf{link}}\rangle,
\end{equation}
where $\Omega=\prod_{u\in V}\frac12 (I+S_u)$ is the projector onto the codespace of the $C4$-code. 
Write $\Omega=\Omega_A\Omega_B$, where 
\[
\Omega_A=\prod_{u\in A}\frac12 (I+S_u) \quad \mbox{and} \quad
\quad \Omega_B=\prod_{u\in B}\frac12 (I+S_u).
\]
Since $\overline{U}_A$ and $\overline{\Pi}^x_A$ commute with $\Omega_A$, one gets
\begin{equation}
\label{memory7}
p^x_t(m_A)=\gamma \langle \phi_{\mathsf{link}} |\overline{U}_A^\dag (\overline{\Pi}^x_A \Omega_A) \overline{U}_A
\Omega_B |\phi_{\mathsf{link}}\rangle.
\end{equation}
Assume first that $B\ne\emptyset$.
Expand the projector $\Omega_B$ as
\[
\Omega_B=2^{-|B|} \sum_{C\subseteq B} S_C, \qquad S_C\equiv \prod_{u\in C} S_u.
\]
Consider a link operator $L_e$ associated with some edge $e$ such that both endpoints
of $e$ belong to $B$. Such a link operator commutes with all operators acting  on $A$.
Note that $L_e$ commutes with all vertex stabilizers $S_u$ except for the two
stabilizers  located at the endpoints of $e$. It follows that 
$L_e S_C L_e=-S_C$ if exactly one end-point of $e$ belongs to $C$.
Furthermore, since $\phi_{\mathsf{link}}$ is stabilized by all link operators one
infers that $\langle \phi_{\mathsf{link}} |O_A S_C |\phi_{\mathsf{link}}\rangle=0$ 
for any operator $O_A$ unless $C=\emptyset$ or $C=B$. 
Here we  used the fact that $B$ is a connected set
for any $t$, see Fig.~\ref{fig:order}. 
Substituting the expansion of $\Omega_B$
into Eq.~(\ref{memory7}) and using the above observation gives
\begin{equation}
\label{memory8}
p^x_t(m_A)=\gamma' \langle \phi_{\mathsf{link}} |\overline{U}_A^\dag (\overline{\Pi}^x_A \Omega_A) \overline{U}_A
(I+S_B) |\phi_{\mathsf{link}}\rangle,
\end{equation}
where $\gamma'=2^{-|B|} \gamma$. 
Next we note that $\phi_{\mathsf{link}}$ is stabilized by $S_A S_B$ since 
the latter coincides with the product of all link operators $L_e$ (including 
$ic_1c_2$ and $ic_3c_4$) and $\phi_{\mathsf{link}}$ is stabilized by any link operator. 
Thus one can replace $S_B$ by $S_A$ in Eq.~(\ref{memory8}). 
However, since $S_A$ commutes with $\overline{U}_A$ and
$S_A\Omega_A=\Omega_A$, we arrive at
\begin{equation}
\label{memory9}
p^x_t(m_A)=2\gamma' \langle \phi_{\mathsf{link}} |\overline{U}_A^\dag (\overline{\Pi}^x_A \Omega_A) \overline{U}_A|\phi_{\mathsf{link}}\rangle.
\end{equation}
Using the identity
\[
\overline{\Pi}^x_A \Omega_A=\prod_{u\in A} \frac12 (I+m_u \overline{X}_u)  \frac12(I+m_u \overline{X}_u S_u)
\]
one finally gets
\begin{equation}
\label{memory10}
p^x_t(m_A)=2^{t+1-n} \gamma \| G_{2t} G_{2t-1} \cdots G_2 G_1 \phi_{\mathsf{link}}\|^2
\end{equation}
where $G_{2a-1}$, $G_{2a}$ are operators acting non-trivially only on the subset of modes $\Gamma_a$,
namely
\[
G_{2a-1}= \frac12(I+m_a \overline{X}_a) e^{i\eta_a \overline{Z}_a},
\]
and
\[
G_{2a}=\frac12(I+m_a \overline{X}_aS_a).
\]
Noting that  $\overline{Z}_a$,  $\overline{X}_a$ and $\overline{X}_a S_a$ have the form 
$ic_p c_q$ for some $p,q\in \Gamma_a$, see Eqs.~(\ref{C4stab},\ref{C4logical}),
one concludes that  Eq.~(\ref{memory10}) includes only FLO gates.

The above arguments also apply to the case $B=\emptyset$, that is, $t=n$.
The only difference is that now Eq.~(\ref{memory8}) has no term $S_B$
and thus Eq.~(\ref{memory10}) becomes
\begin{equation}
\label{memory10final}
p^x(m)=\gamma \| G_{2n} G_{2n-1} \cdots G_2 G_1 \phi_{\mathsf{link}}\|^2.
\end{equation}

Let us discuss the cost of sampling $m$ from $p^x(m)$.
First,  observe that any single-qubit marginal state of $\psi_L$ is maximally mixed
since the surface code has no stabilizers of weight one. 
Since the same is true about the state $U|\psi_L\rangle$, one can
pick the first measurement outcome $m_1$ at random from the uniform distribution.
Suppose we have already sampled $m_1,\ldots,m_{t-1}$
and the simulator's current state is 
\begin{equation}
\label{phi_t}
|\phi_{t-1}\rangle =G_{2t-2} G_{2t-3}\cdots G_2 G_1 |\phi_{\mathsf{link}}\rangle.
\end{equation}
Plugging Eqs.~(\ref{memory10},\ref{memory10final}) into Eq.~(\ref{conditional}) gives
\begin{equation}
\label{conditional1}
p^x_t(m_t|m_1,\ldots,m_{t-1}) =\frac{\kappa_t \| G_{2t} G_{2t-1}\phi_{t-1} \|^2}{\|\phi_{t-1}\|^2},
\end{equation}
where $\kappa_t=2$ for $t<n$ and $\kappa_n=1$.
The conditional probability of the outcome $m_t=1$ can be computed
by simulating two more FLO gates $G_{2t}$, $G_{2t-1}$
starting from the state $\phi_{t-1}$
which takes time $O(n^2)$. Once the conditional probability is computed,
one can sample $m_t$ by tossing a suitably biased coin. 
This produces the next syndrome $m_t$ together with the next state $\phi_t$.
After $n$ iterations one gets the desired sample $m$ from $p^x(m)$. 
Since the above algorithm uses $O(n)$ FLO gates, the simulation runtime scales
as $O(n^3)$. 

As before, we can reduce the runtime to $O(n^2)$ by 
exploiting the fact that the initial state $\phi_{\mathsf{link}}$ has a product form
and by observing that once a pair of modes have been measured, it can be 
removed from the simulator. 
Indeed, consider some intermediate step $t$
and let $j$ be the column 
of the lattice that contains $t$-th qubit. Then all modes in the columns
$j+2,\ldots,d$ can be grouped into unentangled pairs located on edges
such that each pair is  stabilized by the respective  link operator $L_e$. 
Such unentangled pairs do not need to be loaded into the simulator.
Likewise,  all modes in the columns $1,\ldots,j-2$ have already been  measured
and can be removed from the simulator. Thus at any given time step
the number of ``active" modes that needs to be simulated is only  $O(n^{1/2})$. 
Accordingly, the cost of simulating a single FLO gate is at most $O(n)$.
Since the number of gates is $O(n)$, the total simulation cost is $O(n^2)$.

{\em Remark~1:} The same reasoning shows that the probability $p^x(m)$
of any given outcome $m$ can be computed up to the normalizing coefficient $\gamma$
in time $O(n^2)$ by simulating the FLO circuit defined in Eq.~(\ref{memory10final}).
Furthermore, the normalizing coefficient $\gamma$ depends only on $n$ (but not on the
rotation angles $\eta_a$).

{\em Remark~2:}  Below we shall also use a slightly modified version of the above algorithm
where the initial state $\psi_L$ is an eigenvector of the logical-$Y$ operator. 
The modified version is exactly the same as above except that the initial state 
$\phi_{\mathsf{link}}$ in Eq.~(\ref{memory5}) is defined as
\begin{equation}
\label{memory5y}
|\phi_{\mathsf{link}}\rangle\langle \phi_{\mathsf{link}}|=\frac12(I-ic_1c_3) \frac12 (I+ic_2c_4) \prod_{e\in E} \frac12 (I+L_e).
\end{equation}
This corresponds to initializing the unpaired modes in the logical-$Y$ state. 

\subsection{Computing the logical rotation angle}
\label{subs:logical}

It remains to show how to compute the logical rotation angle $\theta_s$ for a given 
syndrome $s$. Let us first initialize the logical qubit in the $X$-basis, that is,
we choose  $|\psi_L\rangle=|+_L\rangle$.
Let $\phi_s$ be the final logical state defined in Eq.~(\ref{final2}). 
Define  logical amplitudes
\begin{equation}
\label{rPlus}
A^+_s=\langle +_L| \phi_s\rangle  = \cos{(\theta_s)}
\end{equation}
and
\begin{equation}
\label{rMinus}
A^{-}_s =\langle +_L|Z_L| \phi_s\rangle  = i\sin{(\theta_s)}.
\end{equation}
Here we used Lemma~\ref{lemma:rotation}.
Then 
\begin{equation}
\label{tau}
\tan^2{(\theta_s)} = 
\left| \frac{A^{-}_s}{A^+_s} \right|^2.
\end{equation} 
Using  Eq.~(\ref{final2}) and the identity $\Pi_s =C_s \Pi_0 C_s$ one gets
\begin{equation}
\label{ttau}
\tan^2{(\theta_s)} =\frac{ |\langle +_L |Z_L C_s U |+_L\rangle|^2}{|\langle +_L|  C_s U |+_L\rangle|^2}.
\end{equation}

Let $U_+=C_s U$ and $U_-=Z_LC_s U$.
By definition, $U_\pm$ are  products of single-qubit $Z$ rotations:
\[
U_\pm = \prod_{u=1}^n e^{i \eta_u^{\pm} Z_u}.
\]
Let us expand the logical state  $|+_L\rangle$  in the $Z$-basis:
\[
|+_L\rangle = |\calL|^{-1/2}  \sum_{x\in \calL} |x\rangle,
\]
where $\calL$ is the set of basis states
$x\in \{0,1\}^n$  that obey $Z$-stabilizers of the surface code
(we do not need an explicit formula for $\calL$).  
Since $U_{\pm}$ is diagonal in the $Z$-basis, 
\begin{equation}
\label{rpm2}
\langle +_L|U_\pm |+_L\rangle=2^{n/2} |\calL|^{-1/2} \langle +^{\otimes n} |U_\pm |+_L\rangle.
\end{equation}
Substituting this into Eq.~(\ref{tau}) gives
\begin{equation}
\label{tau1}
\tan^2{(\theta_s)} =\frac{ | \langle +^{\otimes n} |U_- |+_L\rangle |^2}{| \langle +^{\otimes n}|U_+ |+_L\rangle |^2}
\equiv \frac{p_-}{p_+}.
\end{equation}
Since $U_\pm$ is a tensor product of $Z$-rotations, $p_\pm$ is a special
case of the probability  $p^x(m)$ defined in Eq.~(\ref{memory1})
with  $m_u=1$ for all $u$. 
We have already shown that one can compute $\gamma^{-1} p_\pm$
in time $O(n^2)$, where $\gamma$ depends only on $n$, 
see Remark~1 at the end of Section~\ref{subs:syndrome}. 
Thus $\tan^2{(\theta_s)}=p_-/p_+$ can be computed in time $O(n^2)$.

Next let us  initialize  the logical qubit in the $Y$-basis state:
$|\psi_L\rangle=|Y_L\rangle$, where $|Y\rangle\equiv (|0\rangle  +i|1\rangle)/\sqrt{2}$.
Define  logical amplitudes
\begin{equation}
\label{rPlusY}
B^+_s=\langle +_L| \phi_s\rangle  =e^{i\pi/4}( \cos{(\theta_s)}+\sin{(\theta_s)})
\end{equation}
and
\begin{equation}
\label{rMinusY}
B^{-}_s =\langle +_L|Z_L| \phi_s\rangle  =e^{-i\pi/4} (\cos{(\theta_s)}-\sin{(\theta_s)}).
\end{equation}
 The same arguments as above show that 
\begin{equation}
\label{tau2}
\tan^2{(\theta_s-\pi/4)} =\frac{ | \langle +^{\otimes n} |U_- |Y_L\rangle |^2}{| \langle +^{\otimes n}|U_+ |Y_L\rangle |^2}
\equiv \frac{q_-}{q_+}.
\end{equation}
Since $U_\pm$ is a product of $Z$-rotations, $q_\pm$ is a special
case of the probability  $p^x(m)$ defined in Eq.~(\ref{memory1})
with  $m_u=1$ for all $u$ and the initial state $\psi_L$ chosen
as a logical $Y$-basis state. We have already shown that one can compute $q_\pm$
in time $O(n^2)$ see Remarks~1,2 at the end of Section~\ref{subs:syndrome}.
Thus $\tan^2{(\theta_s-\pi/4)}=q_-/q_+$ is computable in time $O(n^2)$. 
Combining Eqs.~(\ref{tau1},\ref{tau2}) gives
\begin{equation}
\label{tantan}
\cos{(2\theta_s)}=\frac{p_+- p_-}{p_+ + p_-}
\quad \mbox{and} \quad 
\sin{(2\theta_s)}=\frac{q_+- q_-}{q_+ + q_-}.
\end{equation}
This determines the logical rotation angle modulo $\pi$.

\section{Numerical results}
\label{sec:numerics}
We implemented the  algorithms described above 
for  translation-invariant coherent noise
and surface codes with distance $5\le d\le 49$.
The smallest distance $d=3$ was skipped because
of strong finite-size effects (note that the 
considered surface codes are only defined for odd values of $d$). 
We used the maximum distance $d=37$  for storage simulations
and $d=49$ for state preparation simulations. 
The logical error rate $P^L$ 
was estimated by the Monte Carlo method with at least $50,000$ syndrome
samples per data point. (The only exception is Fig.~\ref{fig:prepXZ}
where we used $5,000$ syndrome samples per data point.)

\subsection{Numerical results for storage}

\begin{figure*}[htp]
\centering
\subfloat[]{\includegraphics[height=7cm]{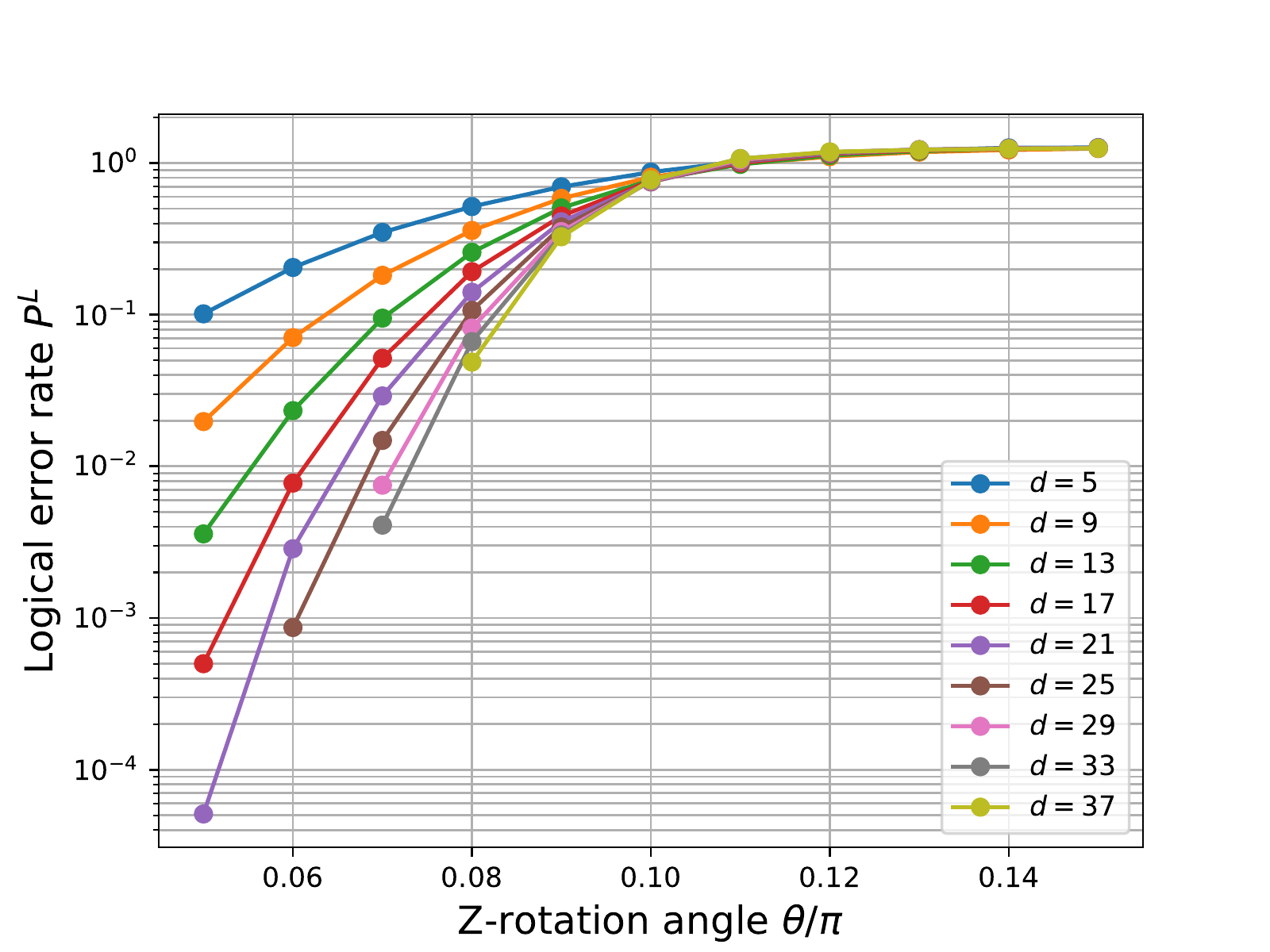}}
\subfloat[]{\includegraphics[height=7cm]{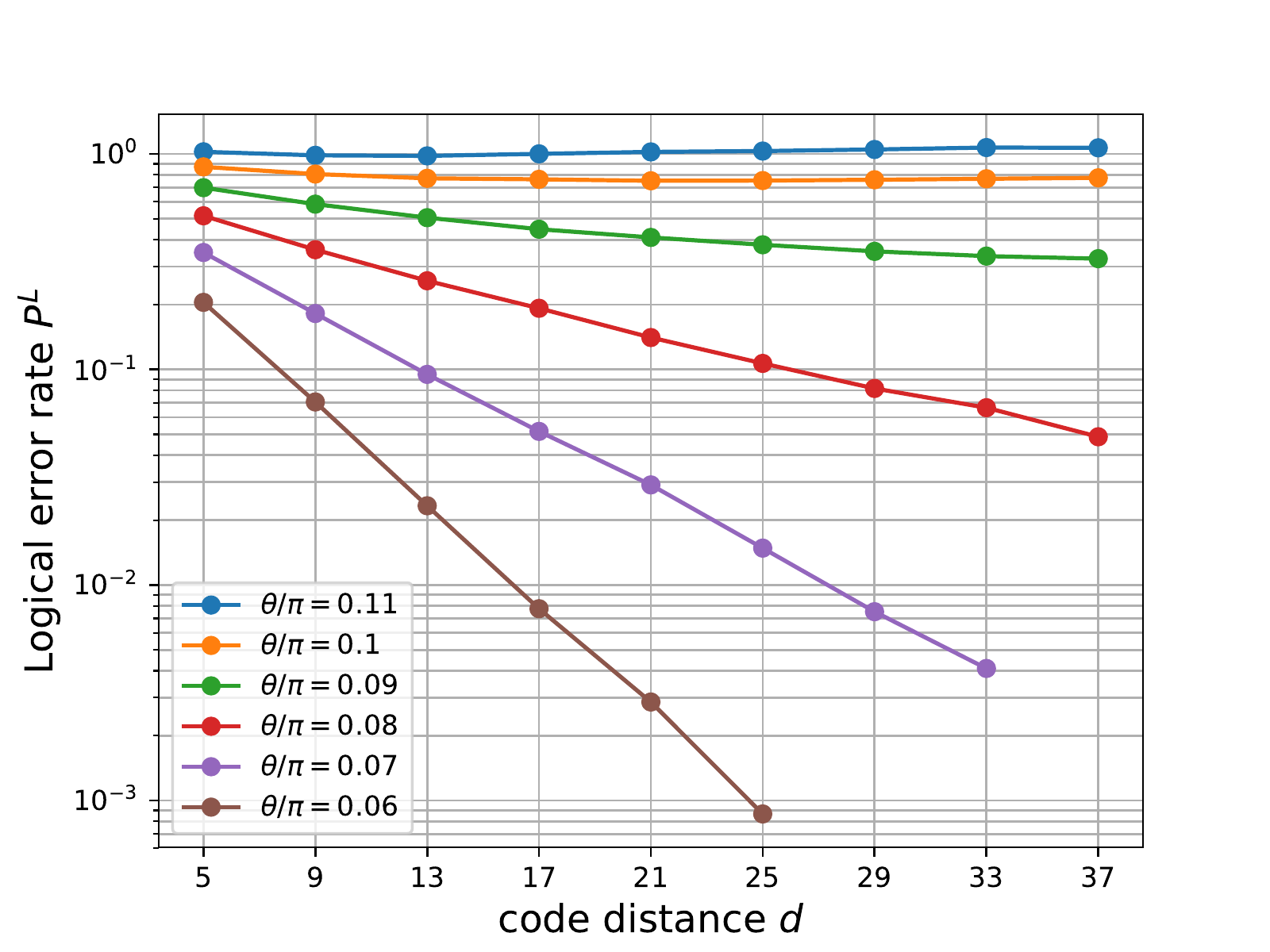}}
\caption{Logical error rate $P^L$  for storage of quantum states.
We consider distance-$d$ surface codes subject to coherent errors $\exp{(i\theta Z)}$ on each qubit.
}
\label{fig:memorybelowthreshold}
\end{figure*}
Consider first the protocol~A for storage of a logical state in the presence of $Z$-rotation errors.
In this section we only consider translation-invariant errors of the form 
$\exp{(i\theta Z)}^{\otimes n}$, where $\theta$ is the only noise parameter. 
Recall that we define the logical error rate as
\begin{equation}
\label{PLrestated}
P^L =2 \sum_s p(s)|\sin \theta_s|,
\end{equation}
where $p(s)$ is the probability of observing a syndrome $s$
and $\theta_s$ is the logical rotation angle conditioned on the syndrome,
see Lemma~\ref{lemma:rotation} in Section~\ref{sec:store}.
To motivate this definition, 
consider  a conditional logical channel
\begin{equation}
\label{LambdaS}
\Lambda_s(\rho)=e^{i\theta _s Z} \rho e^{-i\theta_s Z}
\end{equation}
that describes the residual logical error
 for a given syndrome $s$.
Let $\| \cdot \|_\diamond$ denote the diamond-norm~\cite{kitaev1997quantum} on the space of quantum
channels and  $\mathsf{id}$ be the single-qubit identity channel. 
The identity $\| \Lambda_s - \mathsf{id}  \|_\diamond =2  |\sin \theta_s|$ shows that
$P^L$ coincides with 
 the average diamond-norm distance between the conditional logical
channel and  the identity channel,
\[
P^L=\sum_s p(s) \| \Lambda_s - \mathsf{id}  \|_\diamond.
\]
Using the symmetries of the surface code one can easily check that $P^L$  is invariant under flipping the sign of $\theta$. Accordingly, it suffices to simulate $\theta \ge 0$.

\begin{figure*}[htp]
\subfloat[]{\includegraphics[height=7cm]{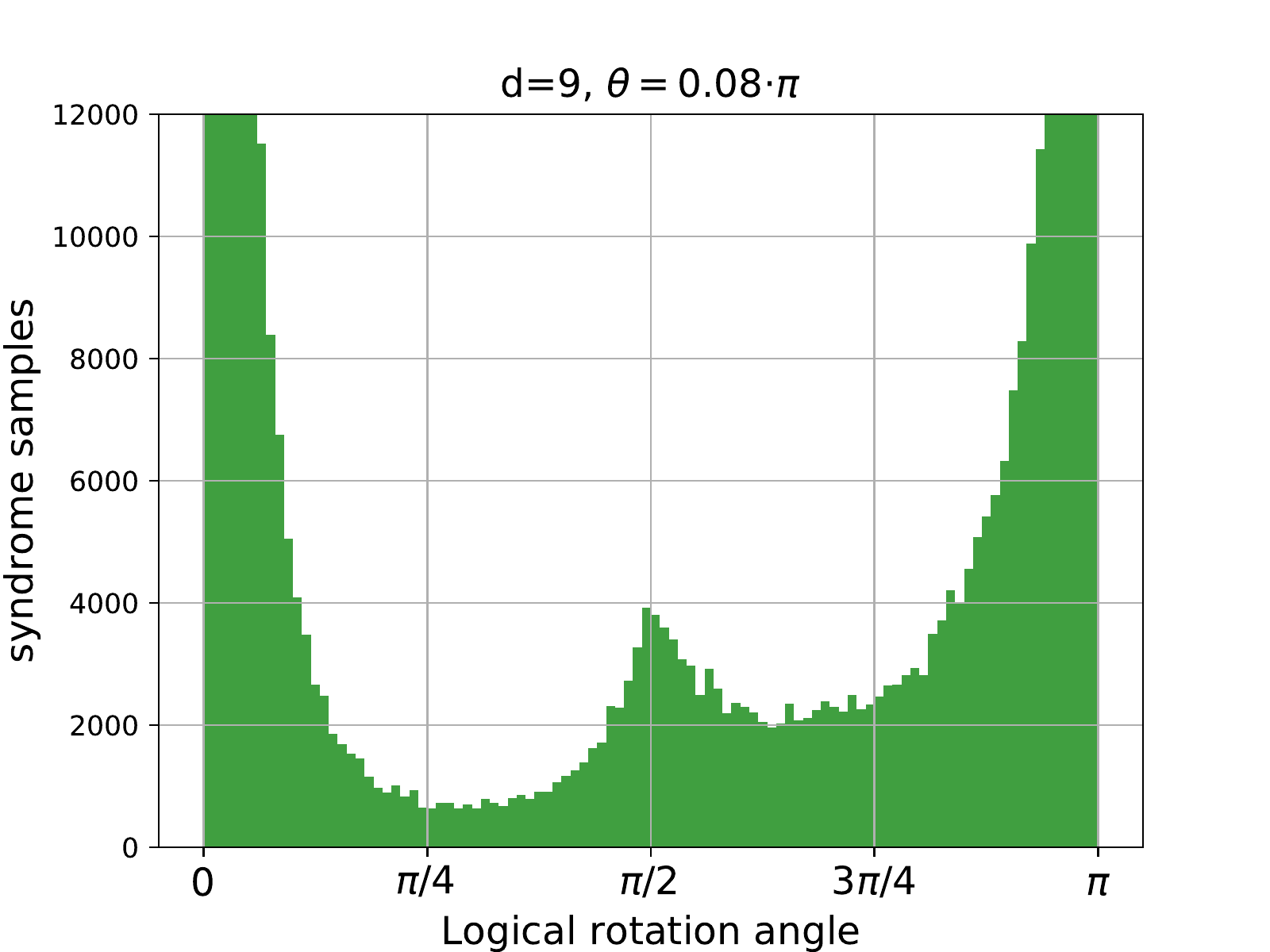}}
\subfloat[]{\includegraphics[height=7cm]{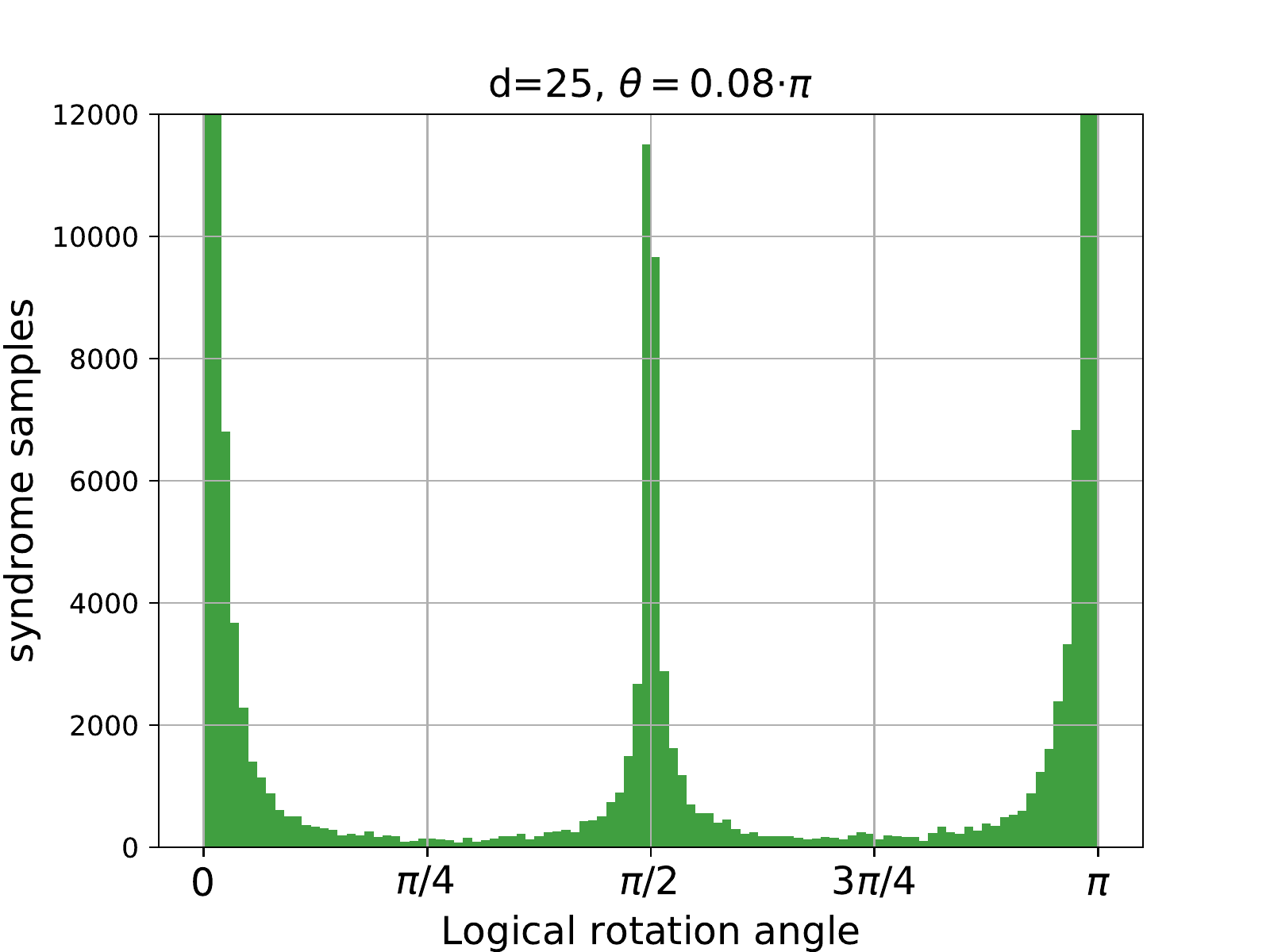}}
\caption{These histograms show the empirical probability distribution of logical rotation angles~$\theta_s$
for the code distance $d=9$ (left) and $d=25$ (right).
The histograms use the same noise parameter $\theta=0.08\pi$.
For ease of visualization, we truncated the main peak at $\theta_s=0$.}
\label{fig:logicalrotationangle}
\end{figure*}

\begin{figure*}[htp]
\subfloat[]{\includegraphics[height=7cm]{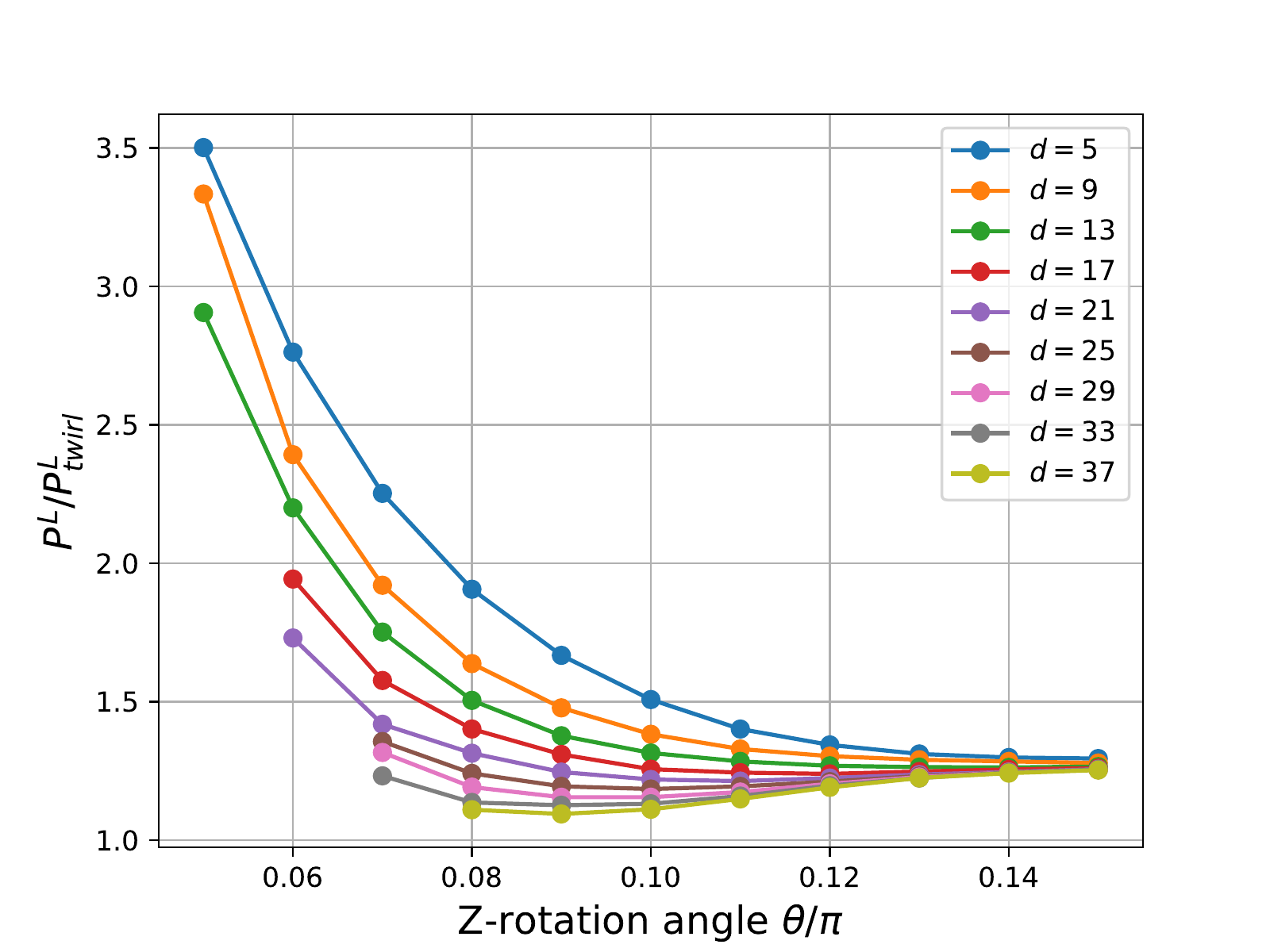}}
\subfloat[]{\includegraphics[height=7cm]{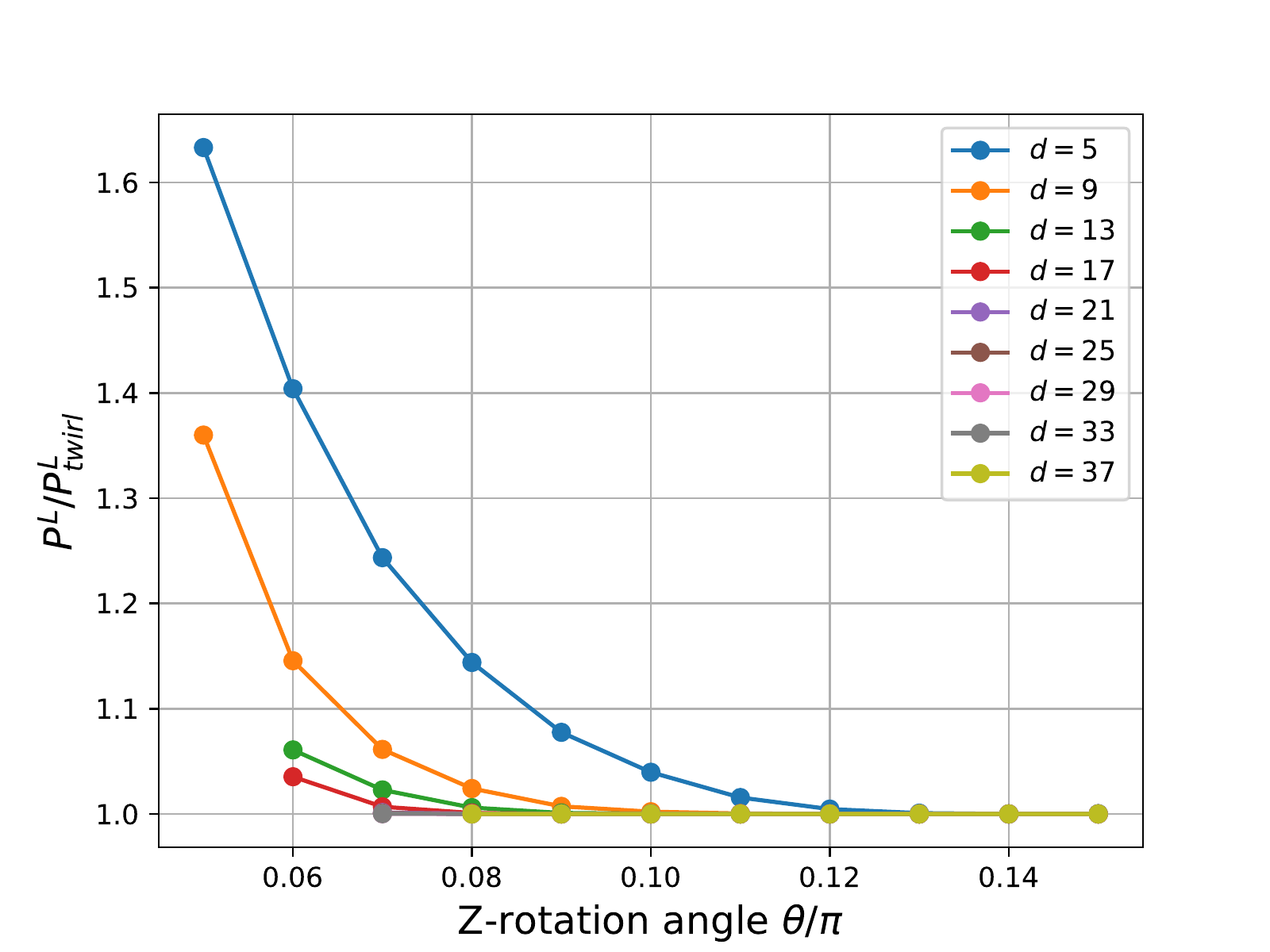}}
\caption{Coherence ratio $P^L/P^L_{\mathsf{twirl}}$ 
for the conditional logical channel (left) and for the average logical channel (right).
In both cases increasing the code distance makes the logical-level noise less coherent. }
\label{fig:memoryincoherence}
\end{figure*}

Our numerical results for the logical error rate
are presented in  Fig.~\ref{fig:memorybelowthreshold}.
The data suggests that the quantity $P^L$ decays exponentially in the code
distance $d$ for $\theta<\theta_0$, where 
\begin{equation}
\label{num:threshold}
0.08\pi \le \theta_0  \le 0.1\pi
\end{equation}
can be viewed as an error correction threshold. 
We observe the exponential decay of $P^L$ as  a function of $d$
in the sub-threshold regime.

Although the logical error rate $P^L$ is a meaningful measure
of how well the initial logical state is preserved, it provides
 no insight into the structure of residual
logical errors.  Algorithm~A 
gives us a unique opportunity to investigate the logical-level noise since it 
outputs both the syndrome $s$ and the 
the logical rotation angle $\theta_s$ conditioned on  $s$.
Fig.~\ref{fig:logicalrotationangle} shows the empirical 
probability distribution of $\theta_s$ obtained by 
sampling $10^6$ syndromes $s$ 
 for the physical $Z$-rotation angle $\theta=0.08\pi$
(which we expect to be slightly below the threshold). We compare the cases $d=9$ and $25$.
In both cases the distribution has a sharp peak at $\theta_s=0$ (equivalent to $\theta_s=\pi$).
This peak indicates that error correction almost always succeeds in the considered regime.
For ease of visualization, we truncated the peak at $\theta_s=0$ on the histograms. 
It can be seen that increasing the code distance has a dramatic effect on the distribution of $\theta_s$.
The distance-$9$ code has a broad
distribution of $\theta_s$ meaning that the logical-level noise
retains  a strong coherence. On the other hand, the distance-$25$ code
has a sharply peaked distribution of $\theta_s$ with a peak at $\theta_s=\pi/2$
which corresponds to the logical Pauli error $Z_L$.
 Such errors are likely to be caused by  ``ambiguous" syndromes $s$ 
for which the minimum weight matching decoder makes a wrong choice
of the Pauli correction $C_s$.
We conclude that as the code distance increases,  the logical-level noise 
can be well approximated by random Pauli errors 
even though the physical-level noise is coherent. 

\begin{figure*}[htp]
\subfloat[]{\includegraphics[height=7cm]{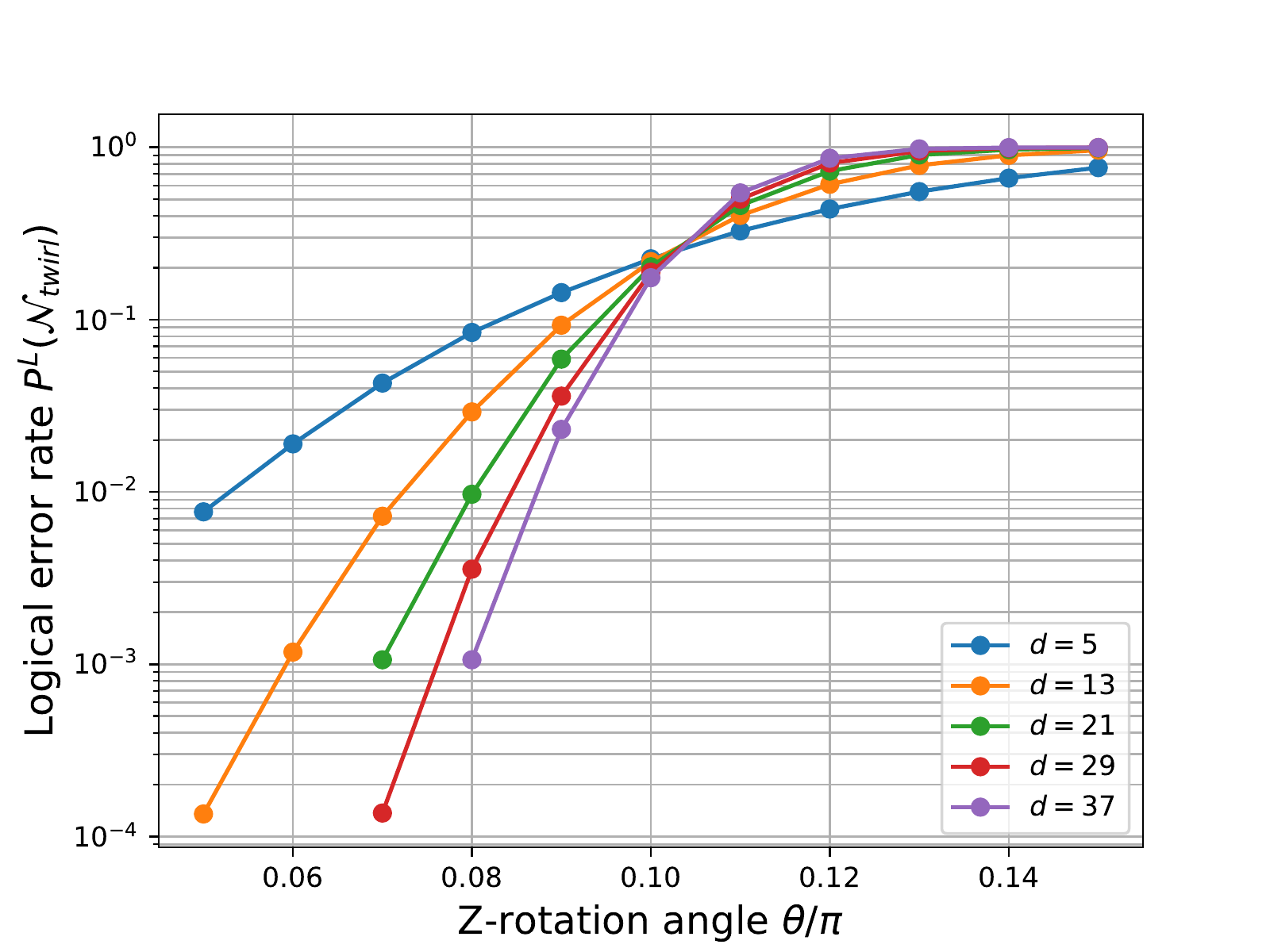}}
\subfloat[]{\includegraphics[height=7cm]{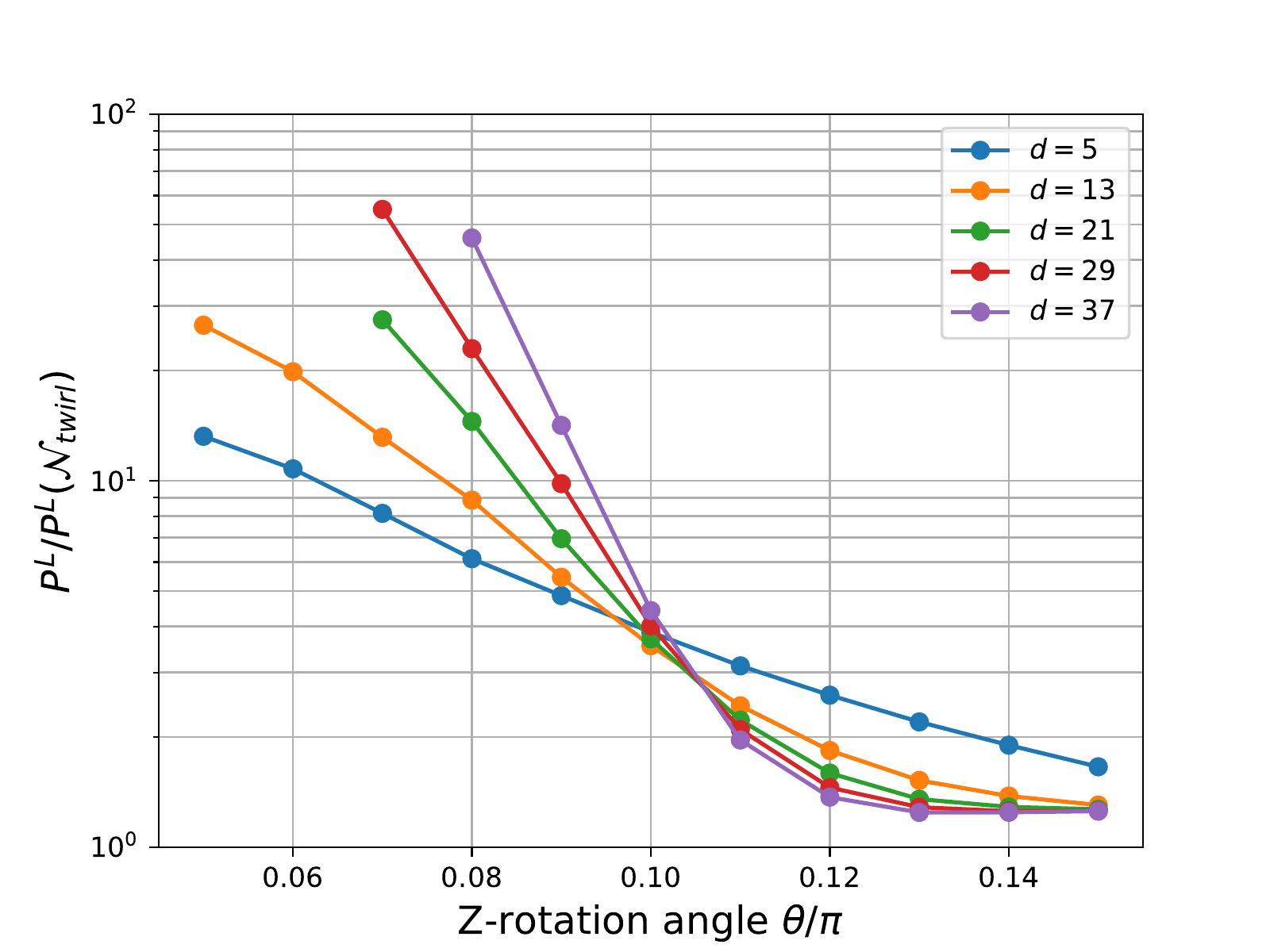}}
\caption{Comparison between the logical error rates $P^L$ and $P^L(\calN_{\mathsf{twirl}})$
computed for 
coherent noise $\calN(\rho)=e^{i\theta Z} \rho e^{-i\theta Z}$
and its Pauli twirled version $\calN_{\mathsf{twirl}}(\rho)=(1-\epsilon) \rho + \epsilon Z\rho Z$ with
$\epsilon=\sin^2{(\theta)}$. In both cases we used the minimum weight matching decoder.
The plot demonstrates that applying the Pauli twirl approximation
to the physical noise significantly underestimates the logical error rate in the sub-threshold regime.}
\label{fig:physical_twirl}
\end{figure*}

To investigate this effect more systematically,  it is desirable to have a metric 
quantifying the degree of coherence present in  the logical-level noise.
To this end let us consider the twirled version of the logical channel $\Lambda_s$,
\[
\Lambda_s^\mathsf{twirl}(\rho)=
 (1-\epsilon_s) \rho + \epsilon_s Z \rho Z, \quad \epsilon_s\equiv \sin^2{(\theta_s)},
\]
and the corresponding logical error rate 
\begin{align}
\label{PLtwirled}
P^L_{\mathsf{twirl}}&=\sum_s p(s) \|\Lambda^{\textsf{twirl}}_s-\mathsf{id}\|_\diamond
=2\sum_s p(s) \sin^2{(\theta_s)}.
\end{align}
Comparison of Eqs.~(\ref{PLrestated},\ref{PLtwirled}) reveals that 
$P^L\ge P^L_{\mathsf{twirl}}$ with the equality 
iff the distribution of $\theta_s$ has all its weight on $\{0,\pi/2\}$, that is,
when the logical noise is incoherent.
It is therefore natural to measure coherence of the logical  noise
by the ratio $P^L/P^L_{\mathsf{twirl}}$.
This ``coherence ratio" is plotted as a function of $\theta$
on Fig.~\ref{fig:memoryincoherence}(a).
The data indicates that the coherence ratio decreases for increasing system size
approaching one for large code distances. This further supports
the conclusion that the logical noise has a negligible coherence. 
Finally, in Fig.~\ref{fig:memoryincoherence}(b), we show the analogous quantity for
the average logical noise channel~\cite{rahn} defined as
\[
\Lambda=\sum_s p(s) \Lambda_s.
\]
This average channel provides an appropriate model for the logical-level noise
if the environment has no access
to the measured syndrome. This may be relevant, for instance, in the quantum communication settings
where noise acts only during transmission of information. 
Thus one can alternatively define the coherence ratio as 
\begin{align}
P^L/P^L_{\mathsf{twirl}}&=\frac{\|\Lambda-\mathsf{id}\|_\diamond}{\|\Lambda^{\mathsf{twirl}}-\mathsf{id}\|_\diamond}\ ,\label{eq:incoherencefract}
\end{align}
where $\Lambda^{\mathsf{twirl}}$ is the Pauli-twirled version of $\Lambda$. 
In our case 
$\Lambda(\rho)=(1-\epsilon) \rho + \epsilon Z\rho Z + i\delta (Z\rho -\rho Z)$,
where $\epsilon=\sum_s p(s) \sin^2{(\theta_s)}$ and $\delta=\sum_s p(s) \sin{(2\theta_s)}/2$,
see Eq.~(\ref{LambdaS}).
A simple calculation yields 
\[
\| \Lambda - \mathsf{id}\|_\diamond = 2\sqrt{\epsilon^2 + \delta^2}
\quad \mbox{and} \quad 
\|\Lambda^{\mathsf{twirl}}-\mathsf{id}\|_\diamond =2\epsilon.
\]
The coherence ratio of the average logical channel
is plotted as a function $\theta$ on Fig.~\ref{fig:memoryincoherence}(b).
It provides a particularly strong evidence that in the limit of large code distances, coherent physical noise gets converted into incoherent logical noise.

Finally, let us compare logical error rates $P^L$ computed for coherent
physical noise $\calN(\rho)=e^{i\theta Z} \rho e^{-i\theta Z}$ and 
its Pauli-twirled version $\calN_{\mathsf{twirl}}(\rho)=(1-\epsilon) \rho + \epsilon Z\rho Z$ with
$\epsilon=\sin^2{(\theta)}$. Applying the Pauli twirl at the physical level amounts to 
ignoring the coherent part of the noise. 
Let $P^L(\calN_{\mathsf{twirl}})$ be the logical error rate 
corresponding to $\calN_{\mathsf{twirl}}$.
The plot of $P^L(\calN_{\mathsf{twirl}})$ and the ratio
$P^L/P^L(\calN_{\mathsf{twirl}})$ are 
shown on Fig.~\ref{fig:physical_twirl}.
It can be seen that applying the Pauli twirl approximation to the physical noise gives an accurate estimate of the error threshold but 
significantly underestimates the logical error probability in the sub-threshold regime. 
We conclude that  coherence of noise may have a profound effect on the performance of large surface codes
in the sub-threshold regime which is particularly important for quantum fault-tolerance.

\subsection{Numerical results for state preparation}
\label{subs:lspnum}

Next consider the protocol~B for preparing the logical basis state
$|+_L\rangle$ by performing syndrome measurements on the initial product state
\[
(\exp{(i\varphi X)} \exp{(i\theta Z)} |+\rangle)^{\otimes n}.
\]
Here $\varphi,\theta\in [0,\pi)$ are noise parameters. The ideal 
protocol  corresponds to  $\theta=0$.
Define the logical error rate $P^L$ as the average
trace-norm distance between the final logical state $|\phi_s\rangle\langle \phi_s|$
and the target state $|+_L\rangle\langle +_L|$. Equivalently, 
\begin{equation}
\label{PLprep}
P^L =2^{1/2} \sum_{s} p(s) \sqrt{1-\langle \phi_s |X_L|\phi_s\rangle}.
\end{equation}
Since the considered noise model generates correlations between 
$X$- and $Z$-syndromes, we opted not to use  the minimum-weight matching decoder
(which treats $X$- and $Z$-syndromes independently).
Instead, we used a simplified decoder that chooses a Pauli correction $C_s$
such that  the final logical state $\phi_s$  always obeys $\langle \phi_s |X_L|\phi_s\rangle\ge 0$.
The simplified decoder is optimal in the sense that it minimizes 
the logical error rate $P^L$ under the constraint that $C_s$ is a Pauli operator.
Note that $0\le P^L\le \sqrt{2}$ for all noise parameters. 
The symmetries of the surface code imply that 
$P^L$, considered as a function of $\theta$ and $\varphi$, is invariant under
transformations 
\[
\theta\gets \theta+\frac{\pi}2, \quad
\varphi \gets \varphi+\frac{\pi}2, \quad
\theta \gets -\theta, \quad
\varphi\gets -\varphi.
\]
Thus  it suffices to simulate the region $0\le \theta,\varphi\le \pi/4$.

\begin{figure}[htp]
\includegraphics[height=7cm]{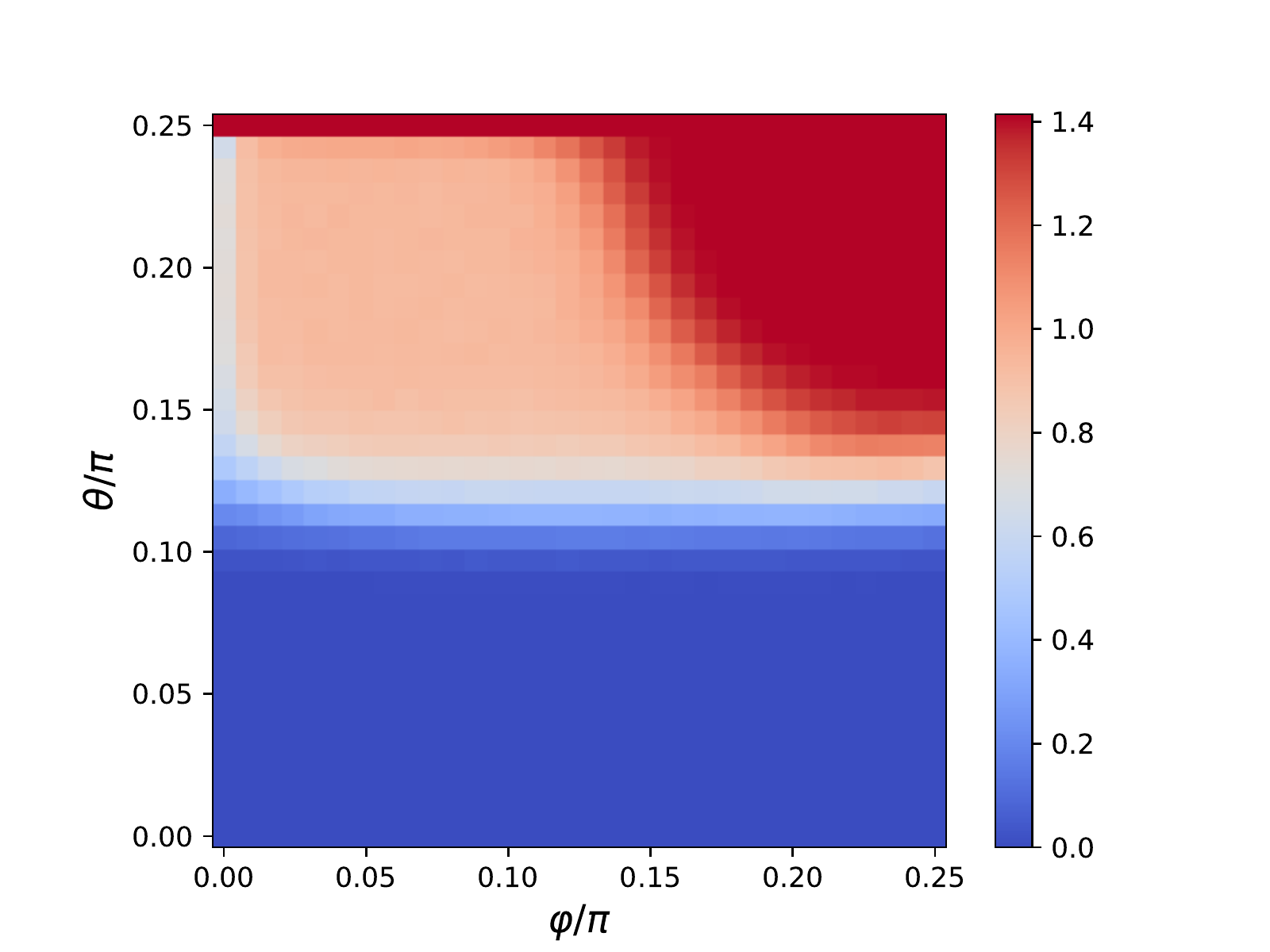}
\caption{Preparation of the  logical   basis state $|+_L\rangle$ 
by performing syndrome measurements on the initial product state 
$(\exp{(i\varphi X)} \exp{(i\theta Z)} |+\rangle)^{\otimes n}$.
The color represents the logical error rate $P^L$ defined in Eq.~(\ref{PLprep}).
The dark blue and red regions represent a ``fault-tolerant" phase where the final 
logical state is close to  $|+_L\rangle$ and $|0_L\rangle$ respectively.
Here we consider a fixed code distance $d=39$.
The plot was generated by sampling 
$5,000$ syndromes  for each pair $(\theta,\varphi)$.
}
\label{fig:prepXZ}
\end{figure}

\begin{figure*}[htp]
\subfloat[]{\includegraphics[height=7cm]{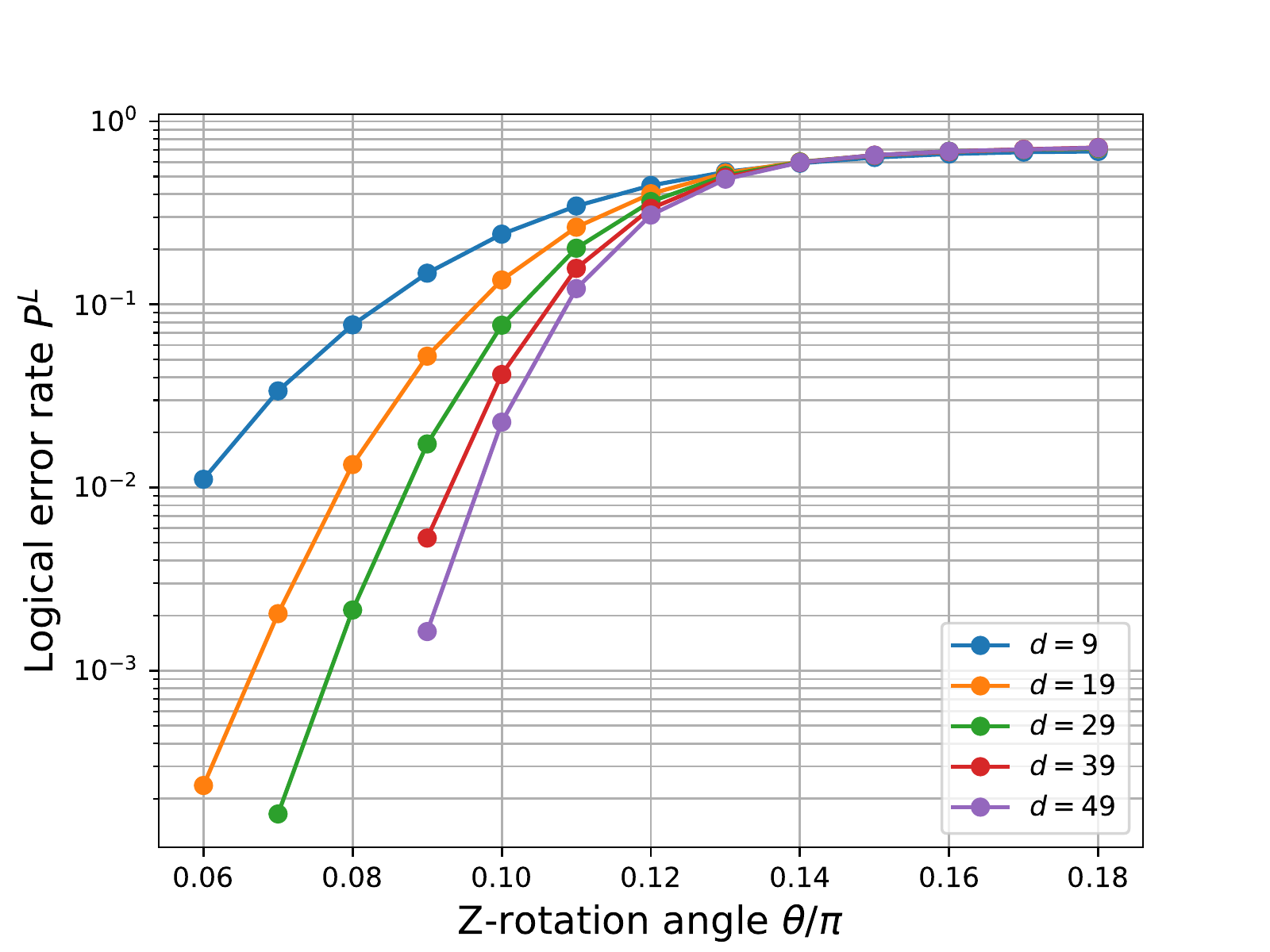}} 
\subfloat[]{\includegraphics[height=7cm]{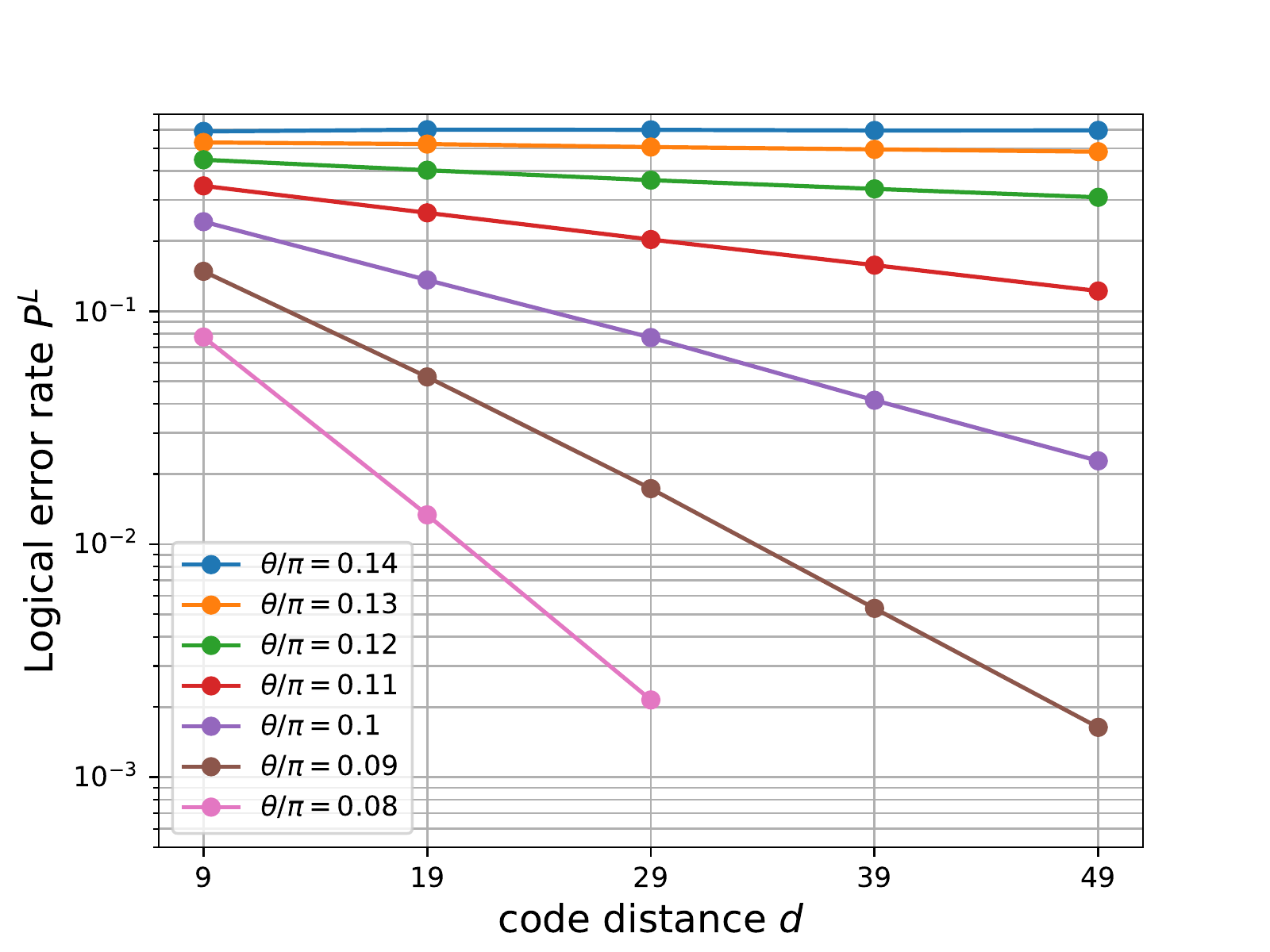}}
\caption{Logical error rate $P^L$ for state preparation 
subject to coherent errors $\exp{(i\theta Z)}$ on each qubit. }
\label{fig:prep}
\end{figure*}

\begin{figure}[t]
\includegraphics[height=7cm]{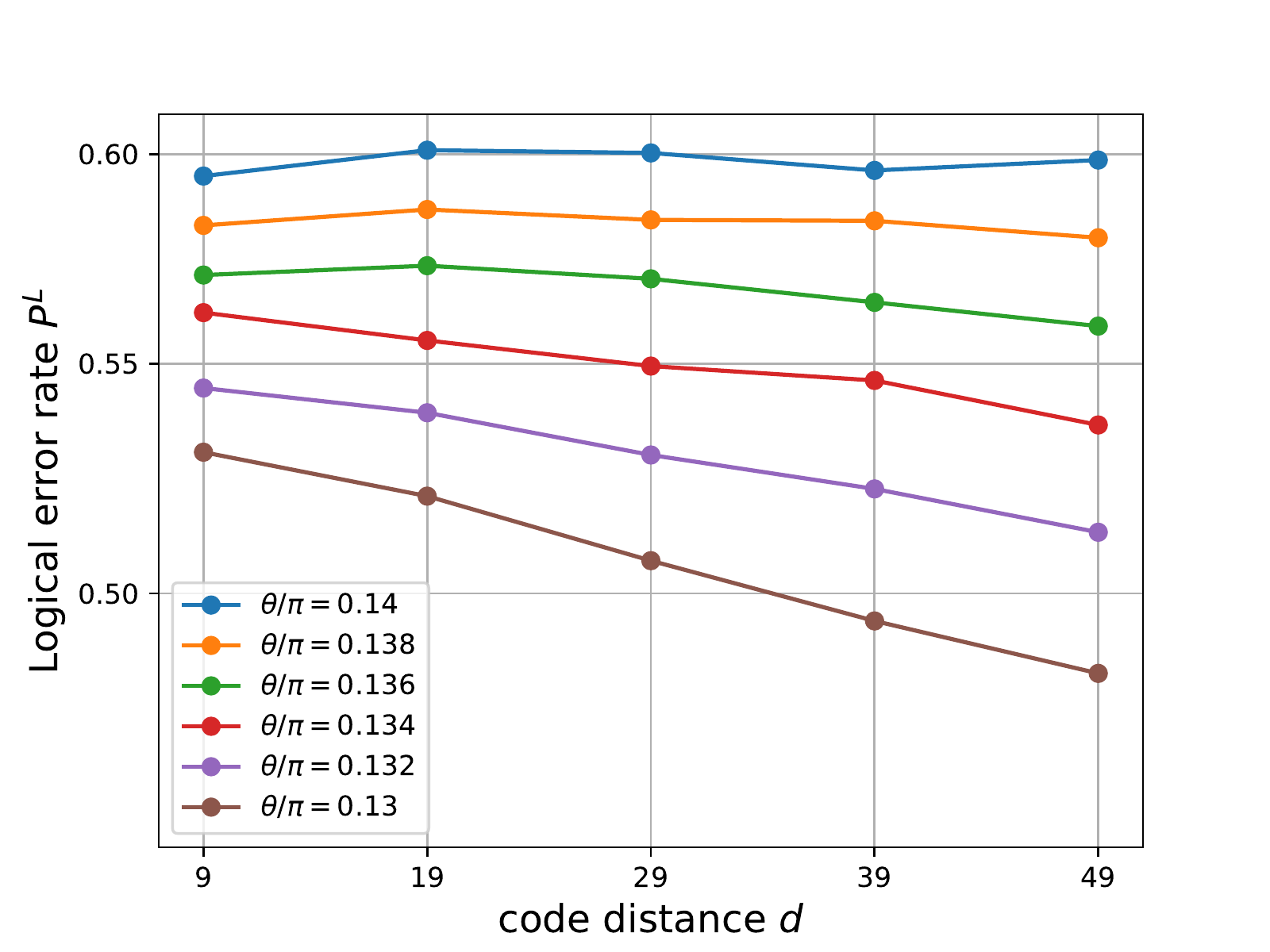}
\caption{Logical error rate $P^L$ for state preparation. Here we only consider angles $\theta$ near the threshold
and $\varphi=0$.}
\label{fig:prep_near}
\end{figure}

Our numerical results for a fixed code distance $d=39$ are presented on Fig.~\ref{fig:prepXZ}.
The data supports  a natural conjecture that the asymptotic behavior of $P^L$ in the limit $n\to \infty$
can be characterized by a single threshold function $\theta_0(\varphi)$ such that 
$\lim_{n\to \infty} P^L=0$ for $\theta<\theta_0(\varphi)$ and $P^L$ is lower bounded
by a positive constant independent of $n$ for $\theta>\theta_0(\varphi)$, see also Fig.~\ref{fig:prep}.
From Fig.~\ref{fig:prepXZ} one gets an estimate
\[
0.1\pi \le \theta_0(\varphi) \le 0.15\pi
\]
for all $\varphi$. The data  indicates that $\theta_0(\varphi)$ has a very mild (if any)
dependence on $\varphi$.
To test the above conjecture, we performed more detailed simulations
for $\varphi=0$, see Figs.~\ref{fig:prep},\ref{fig:prep_near},
obtaining a more refined estimate 
\[
0.13\pi \le \theta_0(0)\le 0.14\pi.
\]

\section{Conclusions}
\label{sec:conclusions}
Our work extends the range of noise models 
efficiently simulable on a classical computer.
It allows -- for the first time -- to numerically investigate the effect of coherent errors  in the regime  of large code sizes
which is important for reliable error threshold estimates. 
 Our simulation algorithms make no assumptions about the particular decoder used. Hence the proposed approach should be universally applicable to  benchmarking the performance of different fault-tolerance strategies in the presence of coherent noise.  

Our numerical results spell good news for quantum engineers pursuing surface code realizations: thresholds for state preparation and storage are reasonably high, suggesting that coherent noise is not
as detrimental as one could expect from the previous studies. 
The numerical investigation of the logical-level noise gives rise to a conceptually appealing conjecture: 
error correction converts coherent physical noise to incoherent logical noise (for large code sizes).
Whether this is an artifact of the considered error correction scheme or manifestation of a more general phenomenon 
is an interesting open question. 

Although we   simulated   only   translation-invariant noise models, all our algorithms
apply to more general qubit-dependent noise. 
This enables numerical study of recently proposed state injection protocols~\cite{lodyga2015simple}, e.g. preparation of
logical magic states, in the presence of coherent errors. 
Another possible application could be testing the so-called  disorder assisted error
correction method~\cite{wootton2011bringing,stark2011localization,bravyi2012disorder}
where  artificial randomness  introduced in the code parameters
suppresses  coherent propagation of errors due to the Anderson localization phenomenon. 
We leave as an open question whether our algorithms can be extended to more general 
coherent noise models such as those including systematic cross-talk errors.

\section*{Acknowledgments}
The authors thank Steven Flammia,  Jay Gambetta, and David Gosset  for helpful discussions. RK is supported by the Technical University of Munich -- Institute for Advanced Study, funded by the German Excellence Initiative and the European Union Seventh Framework
Programme under grant agreement no.~291763. He acknowledges partial support by the National Science Foundation under Grant No.~NSF PHY-1125915, and thanks the Kavli Institute for Theoretical Physics for their hospitality. 
NP acknowledges support by TUM-PREP.
SB acknowledges support from the IBM Research Frontiers Institute
and from Intelligence Advanced Research Projects Activity (IARPA) under contract W911NF-16-0114.

\appendix

\section{Proof of Lemma~\ref{lemma:rotation}}
\label{app:rotation}

We have $C_s=Z(h_s)$ for some $h_s\in \{0,1\}^n$. 
Using the identities $\Pi_s= C_s \Pi_0 C_s$  and $\Pi_0 \psi_L=\psi_L$ one gets
\begin{equation}
\label{ps1}
p(s)= \| \Pi_0 C_s U\Pi_0 \psi_L \|^2.
\end{equation}
Expanding the error $U$ in the Pauli basis gives
\[
U = \sum_{g\in \{0,1\}^n}  \alpha_g i^{|g|}  Z(g)
\]
for some real coefficients $\alpha_g$. Here $|g|$ denotes the Hamming weight 
of a string $g$. Thus 
\[
\Pi_0 C_s U \Pi_0 = \sum_{g\in \{0,1\}^n}
\alpha_g i^{|g|}   \Pi_0 Z(g \oplus h_s) \Pi_0.
\]
Note that $\Pi_0 Z(f) \Pi_0 =0$ unless $Z(f)$ or $Z_L Z(f)$ is a stabilizer
of the surface code. Let $\calA \subseteq \FF_2^n$ be a linear subspace
spanned by $Z$-stabilizers (considered as binary vectors)
and let $Z_L=Z(l)$ for some $l\in \{0,1\}^n$. Then
\[
\Pi_0 C_s U \Pi_0 = a_s \Pi_0 + b_s \Pi_0 Z_L,
\]
where
\[
a_s=\sum_{g\in \calA \oplus h_s} \alpha_g i^{|g|}
\quad \mbox{and} \quad
b_s=\sum_{g\in \calA \oplus h_s \oplus l}  \alpha_g i^{|g|}.
\]
Here $\oplus$ denotes addition of binary strings modulo two. 
Suppose first that $h_s$ has even weight. Since 
any element of $\calA$ has even weight, the sum that defines $a_s$
runs over even-weight vectors $g$, that is, $a_s$ is real. 
Likewise,  since $l$ has odd weight, the sum that defines $b_s$ runs 
over odd-weight vectors $g$, that is, $b_s$ is imaginary. 
Define $q_s=|a_s|^2 + |b_s|^2$. Note that $q_s$ does not depend on $\psi_L$.
One arrives at
\begin{equation}
\label{P0P0}
\Pi_0 C_s U \Pi_0  =\sqrt{q_s} \cdot \Pi_0  \exp{\left[ i \theta_s Z_L \right]},
\end{equation}
where $\theta_s\in [0,2\pi)$ is chosen such that $a_s+ b_s=\sqrt{q_s} e^{i\theta_s}$.
Since $\exp{(i\pi Z_L)}=-I$ and overall phase factors do not matter, one can assume 
$\theta_s\in [0,\pi)$. 
Substituting Eq.~(\ref{P0P0}) into Eq.~(\ref{ps1}) gives
$p(s)=q_s$, that is, $p(s)$ does not depend on $\psi_L$, as claimed. 
Substituting $\Pi_s=C_s \Pi_0 C_s$ into Eq.~(\ref{final2}), noting that $C_s^2=I$,
and using Eq.~(\ref{P0P0}) with $q_s=p(s)$ proves Eq.~(\ref{syn2}).

The case when $h_s$ has odd weight is completely analogous,
except that now $b_s$ is real and $a_s$ is imaginary.
Choose $\theta_s\in [0,2\pi)$ such that $a_s+b_s=i\sqrt{q_s} e^{i\theta_s}$.
Then $\theta_s$ obeys Eq.~(\ref{P0P0}) with an extra factor of $i$ on the right-hand side.
The rest of the proof is exactly as above.

\section{Fermionic linear optics}
\label{sec:gaussian}
 
In this appendix we state some necessary facts on simulation of fermionic linear optics.
The material of this section is  based on 
Refs.~\cite{terhal2002classical,bravyi2004lagrangian,bravyi2011classical}.
Let $\calP_n$ be the group generated by single-qubit Pauli operators
$X_j,Y_j,Z_j$ with $j=1,\ldots,n$. Define Majorana operators
$c_1,\ldots,c_{2n}\in \calP_n$ such that  $c_1=Y_1$, $c_2=X_1$,
\begin{equation}
\label{maj1}
c_{2j-1} =Z_1\cdots Z_{j-1} Y_j \quad \mbox{and} \quad c_{2j} = Z_1\cdots Z_{j-1} X_j
\end{equation}
for $2\le j\le n$. 
They obey commutation rules
\begin{equation}
\label{MCR}
c_p c_q = - c_q c_p \quad \mbox{for $p\ne q$}.
\end{equation}
More generally, given a bit string $x\in \{0,1\}^{2n}$, define
a Majorana monomial $c(x)=c_1^{x_1} c_2^{x_2} \cdots c_{2n}^{x_{2n}}$.
Then 
\begin{equation}
\label{MCR1}
c(x) c(y)=(-1)^{|x\cdot y|} c(y) c(x)
\end{equation}
whenever at least one of the strings $x$,$y$ has even weight. 
Any $n$-qubit operator can be uniquely expressed as a linear 
combination of  the $4^n$ Majorana monomials $c(x)$.

Suppose $\rho$ is a (mixed) $n$-qubit state. 
The {\em covariance matrix} of $\rho$ is a real anti-symmetric matrix
$M$ of size $2n\times 2n$ defined by
\begin{equation}
\label{cov1}
M_{p,q}=\left\{ \begin{array}{rcl} 
\mathrm{Tr}( ic_p c_q \rho)  &\mbox{if} & p\ne q\\
0 &\mbox{if} & p= q\\
\end{array}\right.
\end{equation}
A state $\rho$ is called {\em Gaussian} iff $\rho$ is a linear combination of
only even-weight Majorana monomials $c(x)$, and  the expectation value
of $c(x)$ on $\rho$ can be computed from the covariance matrix $M$ using Wick's theorem.
For example, we require that
\begin{equation}
\label{Wick}
-\mathrm{Tr}( c_p c_q c_r c_s  \rho ) =M_{p,q} M_{r,s}  - M_{p,r} M_{q,s} +  M_{p,s} M_{q,r}
\end{equation}
for all $p\ne q\ne r\ne s$. More generally,  we require that
\begin{equation}
\label{WickGeneral}
\mathrm{Tr}( i^{|x|/2} c(x) \rho ) =\mathrm{Pf}(M[x])
\end{equation}
for all even-weight $x\in \{0,1\}^{2n}$,
where $M[x]$ is a submatrix of $M$ including only rows and columns $p$
with $x_p=1$ and $\mathrm{Pf}$ denotes the Pfaffian.
In the present paper we  only use the special case Eq.~(\ref{Wick}).
To summarize, a Gaussian state can be fully specified by its covariance matrix.

It is well known that FLO gates defined in Section~\ref{sec:methods}
preserve the class of Gaussian states. Thus  a quantum circuit
composed of FLO gates acting on some initial Gaussian state can be  efficiently simulated
if we know how to update the covariance matrix under the action of each gate.
These update rules are stated below.

Consider a pair of modes $p,q\in [1,2n]$ and let 
\begin{equation}
\label{parity}
\Lambda=(1/2)(I+ic_pc_q).
\end{equation}
Note that $\Lambda$ is a projector. It describes 
a post-selective parity measurement for the pair of modes $p,q$
with  the measurement outcome $+1$ (note that the outcome $-1$
can be obtained simply by exchanging $p$ and $q$). 
\begin{fact} 
\label{fact:g1}
If $\rho$ is a Gaussian state with  covariance matrix $M$
then  $\rho'=\Lambda \rho \Lambda/\mathrm{Tr}(\Lambda \rho)$
is a Gaussian state with a covariance matrix $M'$ that can be computed
in time $O(n^2)$ by the following algorithm:
\end{fact}
\begin{center}
\fbox{\parbox{0.9\linewidth}{
\begin{algorithmic}
\Function{MEASURE}{$M,p,q$} 
\State{$\lambda \gets (1/2)(1+M_{p,q} )$}
 \State{\Comment{probability of the outcome $+1$}}
\If {$\lambda\ne 0$} 
\State{$K \gets \mbox{$p$-th column of $M$}$}
\State{$L \gets \mbox{$q$-th column of $M$}$}
\State{$M'\gets M+ (2\lambda)^{-1} ( K L^T - L K^T )$}
\State{Set to zero rows and columns $p,q$ of $M'$}
\State{$M_{p,q}'\gets 1$}
\State{$M_{q,p}'\gets -1$}
\State{\Return $(\lambda,M')$}
\EndIf
\EndFunction
\end{algorithmic}
}}
\end{center}
We note  that the final matrix $M'$ has a block structure such that the modes
$p,q$ are not coupled to any other modes. Thus one can  remove
rows and columns $p,q$ from $M'$ without losing any information.  

Next let us discuss two-mode rotations
\begin{equation}
\label{rotate}
U=\exp{(\gamma c_p c_q)}.
\end{equation}
Here $\gamma\in [0,\pi)$ is the rotation angle. 
One can check that $U$  is a unitary operator
such that the action of $U$ in the Heisenberg picture is
\begin{equation}
\label{rotate1}
\begin{array}{rcl}
U^\dag c_p U  &=& \cos{(2\gamma)} c_p + \sin{(2\gamma)}  c_q, \\
U^\dag c_q U &=& -\sin{(2\gamma)} c_p  + \cos{(2\gamma)} c_q, \\
U^\dag c_r U  & =&c_r, \quad \mbox{if $r\notin \{p,q\}$ }.\\
\end{array}
\end{equation}
We shall describe the transformation Eq.~(\ref{rotate1}) by an orthogonal matrix
$R\in SO(2n)$ such that 
\[
U^\dag c_t U =\sum_{s=1}^{2n} R_{t,s} c_s, \qquad 1\le t\le 2n.
\]
\begin{fact} 
\label{fact:g2}
If $\rho$ is a Gaussian state with a covariance matrix $M$ then $\rho'=U\rho U^\dag$
is a Gaussian state with a covariance matrix $M'=RMR^T$
that can be computed in time $O(n)$.
\end{fact}
Finally, the class of Gaussian states is closed under the tensor product
operation.
\begin{fact}
\label{fact:g3}
Let $\rho_i$ be a  Gaussian state of $n_i$ qubits with a covariance matrix
$M_i$. Here $i=1,2$. 
Then $\rho_1\otimes \rho_2$ is a Gaussian state of $n_1+n_2$ qubits
with a covariance matrix 
\begin{equation}
\label{M1M2}
M_1\oplus M_2 \equiv \left[ \begin{array}{cc}
M_1 & 0 \\
0 & M_2 \\
\end{array}
\right].
\end{equation}
\end{fact}
All our algorithms work by simulating a sequence of two-mode parity measurements
and two-mode rotations starting from a certain simple initial Gaussian state. 
To describe these initial states we need two more facts. 
\begin{fact}
\label{fact:g4}
Let  $\psi$ is a pure single-qubit state with a Bloch vector
$\vec{b}=(b^x,b^y,b^z)$. 
Let $\overline{\psi}$ be the $C4$-encoding of $\psi$
defined in Eqs.~(\ref{C4stab},\ref{C4logical}).
Then $\overline{\psi}$ is Gaussian with a covariance matrix
\begin{equation}
\label{M4}
M=\left[ \begin{array}{cccc}
0 & b^x & -b^y & b^z \\
-b^x & 0 & b^z & b^y \\
b^y & -b^z & 0 & b^x \\
-b^z & -b^y & -b^x & 0\\
\end{array}
\right].
\end{equation}
\end{fact}
As a corollary, any tensor product of single-qubit states encoded by the $C4$-code
is a Gaussian state whose covariance matrix is a direct sum of $4\times 4$
matrices defined in Eq.~(\ref{M4}).

By analogy with stabilizer states, certain Gaussian states can be defined by
an abelian group of Pauli stabilizers. Namely, suppose 
$\sigma \, : \, [2n] \to [2n]$ is a permutation.
\begin{fact}
\label{fact:g5}
There exists a unique  Gaussian state $\rho=|\phi\rangle\langle\phi|$ such that
$ic_{\sigma(2j-1)} c_{\sigma(2j)}|\phi\rangle=|\phi\rangle$ for all $j=1,\ldots,n$.
The state $\rho$ has a covariance matrix
\begin{equation}
\label{Mpaired}
M_{r,s} = \sum_{j=1}^n \delta_{r,\sigma(2j-1)} \delta_{s,\sigma(2j)} - \delta_{r,\sigma(2j)} \delta_{s,\sigma(2j-1)}.
\end{equation}
\end{fact}
The above facts are sufficient to simulate any quantum circuit composed of FLO gates, as
defined in Section~\ref{sec:methods}, and provide all details necessary for implementation
of our algorithms~$A$ and $B$. 

\bibliographystyle{apsrev4-1}

%


\end{document}